\documentclass[]{elsart}

\newcommand{\fraca}{2.0\%}
\newcommand{\fracb}{5.1\%}
\newcommand{\fracc}{31\%}

\newcommand{\Sfracaa}{1.8} % alpha = 1.7
\newcommand{\Sfracba}{5.0} % alpha = 1.7
\newcommand{\Sfracca}{31}  % alpha = 1.7
\newcommand{\Sfracab}{2.0} % alpha = 2.0
\newcommand{\Sfracbb}{5.1} % alpha = 2.0
\newcommand{\Sfraccb}{31}  % alpha = 2.0
\newcommand{\Sfracac}{2.5} % alpha = 2.5
\newcommand{\Sfracbc}{5.4} % alpha = 2.5
\newcommand{\Sfraccc}{31}  % alpha = 2.5
\newcommand{\Sfracad}{2.9} % alpha = 3.0
\newcommand{\Sfracbd}{5.9} % alpha = 3.0
\newcommand{\Sfraccd}{32}  % alpha = 3.0

\newcommand{\fluxa}{$3.8 \times 10^{-3}$}
\newcommand{\fluxb}{$2.5 \times 10^{-3}$}
\newcommand{\fluxc}{$2.2 \times 10^{-3}$}
\newcommand{\fluxunit}{km$^{-2}$~sr$^{-1}$~yr$^{-1}$}

\newcommand{\Sfluxaa}{3.3} % alpha = 1.7
\newcommand{\Sfluxba}{2.4} % alpha = 1.7
\newcommand{\Sfluxca}{2.2} % alpha = 1.7
\newcommand{\Sfluxab}{3.8} % alpha = 2.0
\newcommand{\Sfluxbb}{2.5} % alpha = 2.0
\newcommand{\Sfluxcb}{2.2} % alpha = 2.0
\newcommand{\Sfluxac}{4.7} % alpha = 2.5
\newcommand{\Sfluxbc}{2.6} % alpha = 2.5
\newcommand{\Sfluxcc}{2.2} % alpha = 2.5
\newcommand{\Sfluxad}{5.5} % alpha = 3.0
\newcommand{\Sfluxbd}{2.8} % alpha = 3.0
\newcommand{\Sfluxcd}{2.2} % alpha = 3.0

\newcommand{\effaa}{0.60} % alpha = 1.7
\newcommand{\effba}{0.83} % alpha = 1.7
\newcommand{\effca}{0.93} % alpha = 1.7
\newcommand{\effa}{0.53}  % alpha = 2.0
\newcommand{\effb}{0.81}  % alpha = 2.0
\newcommand{\effc}{0.92}  % alpha = 2.0
\newcommand{\effac}{0.43} % alpha = 2.5
\newcommand{\effbc}{0.76} % alpha = 2.5
\newcommand{\effcc}{0.91} % alpha = 2.5
\newcommand{\effad}{0.36} % alpha = 3.0
\newcommand{\effbd}{0.71} % alpha = 3.0
\newcommand{\effcd}{0.90} % alpha = 3.0

\newcommand{\Ntesta}{2761}
\newcommand{\Ntestb}{1329}
\newcommand{\Ntestc}{372}

\newcommand{\Nnona}{570}
\newcommand{\Nnonb}{145}
\newcommand{\Nnonc}{21}

\newcommand{\Nphoa}{0}
\newcommand{\Nphob}{0}
\newcommand{\Nphoc}{0}

\newcommand{\Lima}{3.0}
\newcommand{\Limb}{3.0}
\newcommand{\Limc}{3.0}

\usepackage{lineno}
\usepackage{graphicx}
\usepackage{epsfig,wrapfig}
\usepackage[utf8]{inputenc}
\usepackage[T1]{fontenc}

\begin{document}

\begin{frontmatter} %necessary for elsart style

\title{
Upper Limit on the Cosmic-Ray Photon Flux Above $10^{19}$~eV Using the
Surface Detector of the Pierre Auger Observatory
}

%\author{The Pierre Auger Collaboration (author list)}

\author{{\bf The Pierre Auger Collaboration:}}
\author{J.~Abraham$^{14}$,} 
\author{P.~Abreu$^{69}$,} 
\author{M.~Aglietta$^{55}$,} 
\author{C.~Aguirre$^{17}$,} 
\author{D.~Allard$^{33}$,} 
\author{I.~Allekotte$^{7}$,} 
\author{J.~Allen$^{89}$,} 
\author{P.~Allison$^{91}$,} 
\author{J.~Alvarez-Mu\~{n}iz$^{76}$,} 
% ...

\begin{abstract}
A method is developed to
search for air showers initiated by photons using data
recorded by the surface detector of the Auger Observatory.
The approach is based on observables sensitive to the longitudinal
shower development, the signal risetime and the curvature of the
shower front.
Applying this method to the data, upper limits on the flux of photons of
\fluxa{}, \fluxb{}, and \fluxc{}~\fluxunit{} above
$10^{19}$~eV, $2\times10^{19}$~eV, and $4\times10^{19}$~eV are derived,
with corresponding limits on the fraction of photons
being \fraca{}, \fracb{}, and \fracc{} (all limits at 95\% c.l.).
These photon limits disfavor certain exotic models of sources of cosmic rays.
The results also show that the approach adopted by the Auger Observatory to
calibrate the shower energy is not strongly biased by a contamination from
photons.
\end{abstract}

\end{frontmatter} % for elsart

%\linenumbers

%%%%%%%%%%%%%%%%%%%%%%%%
% SECTION INTRODUCTION %
\section{Introduction}

The search for photons in the ultra-high energy (UHE)
cosmic-ray flux has been stimulated by the observation of cosmic
rays with energies exceeding 
$E_{\rm GZK} \sim$ 6$\times$10$^{19}$~eV
\cite{linsley,hp_high,flyseye_high,agasa_high,hires-gzk,pao_topten}.
If these particles are due to cosmologically distant sources,
the flux spectrum is expected to steepen above this energy.
Intriguingly, a flux spectrum with no apparent steepening above $E_{\rm GZK}$
has been reported by the AGASA Collaboration~\cite{agasa_spectrum}.
To account for this observation and to circumvent the theoretical
challenge of explaining particle acceleration to such energies,
models involving new physics have been proposed in which the cosmic rays are
created at the observed energies at relatively close distances from the
Earth.
These ``top-down'' models~\cite{td_sigl,sarkar} may involve super heavy
dark matter (SHDM) ~\cite{shdm,aloisio04,ellis}, topological defects~\cite{td},
or neutrino interactions with the relic neutrino background
(Z-bursts)~\cite{zb}.
A common feature of these models is the prediction of a substantial
photon flux at highest energies.

The Auger Collaboration has recently reported a measurement of the
cosmic-ray spectrum from the Auger South site showing a flux suppression
above $E_{\rm GZK}$~\cite{Roth_2007in}.
The Auger method is based on a large surface array to
collect the required statistics and a fluorescence detector to
calibrate the energy scale. Using this ``hybrid'' approach,
the energy reconstruction is largely
independent of hadronic interaction parameters and, in case of
nuclear primaries, of the primary mass composition.
However, as explained later, the energy assignment from surface arrays
can be substantially 
altered in the case of primary photons. This would affect the
reconstructed primary spectrum if a non-negligible number of the
highest-energy
events, where data from the fluorescence telescopes are sparse due to
their $\sim$10\% duty cylce, was
actually due to photons (see also \cite{busca}). 
It is worthwhile to note that the acceptance of fluorescence
detectors (as also applied in the HiRes experiment~\cite{hires-gzk})
can be altered in the case of photon primaries~\cite{desouza,chou,fd_limit}.

UHE photons can also act as tracers of the GZK
(Greisen-Zatsepin-Kuzmin) process~\cite{gzk} of resonant
photopion production of nucleons off the cosmic microwave background.
The corresponding photon fluxes are sensitive to source features
(type of primary, injection spectrum, distance to sources ...) and to
propagation parameters (extragalactic radio backgrounds and magnetic
fields)~\cite{sarkar,gelmini,sigl95,gzk-photon1,gzk-photon2}.

Thus, the search for primary photons remains an important subject for
various reasons~\cite{review}, particularly
\begin{itemize}
\item to set significant limits to the possible contribution of top-down
      mechanisms to the primary cosmic-ray flux;
\item to search for GZK photons, to prove the GZK effect and constrain
      source and propagation models;
\item to establish the maximum photon fraction in the primary flux, for which
      the energy estimate in the surface array detector would be altered;
\item to obtain input to fundamental physics, for instance, to probe
      quantum gravity effects in the electromagnetic
      sector~\cite{galaverni}.
\end{itemize}

Showers initiated by UHE photons develop differently from showers induced
by nuclear primaries.
Particularly, observables related to the development stage or ``age''
of a shower (such as the depth of shower maximum $X_{\rm max})$
and to the content of shower muons provide good sensitivity to identify
primary photons.
Photon showers are expected to develop deeper in the atmosphere
(larger $X_{\rm max}$). This is connected to the smaller multiplicity
in electromagnetic interactions compared to hadronic ones, such that
a larger number of interactions is required to degrade the energy
to the critical energy where the cascading process stops.
Additionally, the LPM effect~\cite{lpm} results in a suppression of the
pair production and bremsstrahlung cross-sections.
Photon showers also contain fewer secondary muons, since
photoproduction and direct muon pair production are expected to play
only a sub-dominant role.

Searches for photons were previously conducted based on surface
arrays~\cite{hp_photon,agasa_photon,risse_photon,ay_limit,yakutsk},
and limits to the fraction of photons were reported
(see~\cite{review} for a review).
The derivation of limits to the photon \emph {fraction} using surface
array data alone is an experimental and conceptual challenge
(see also Section~\ref{sec_energy}).
Firstly, for conclusions on the fraction, the energy scales for photon \emph {and}
nuclear primaries are needed. These energy scales may differ from each
other for surface arrays, and the difference between
the scales may depend in a non-trivial way on primary parameters such as the
shower zenith angle.
Secondly, the energy reconstruction of nuclear primaries suffers from 
substantial uncertainties due to our limited knowledge of high-energy
hadron dynamics.

Both issues can be resolved using the fluorescence technique, which
is near-calorimetric and largely independent of simulating hadron
interactions.
A corresponding approach has been developed and applied recently 
to obtain a first bound on the fraction of photons from data taken
at the Auger Observatory~\cite{fd_limit}.

In this work, using the larger number of events recorded by the
surface array, we derive for the first time a direct limit to the
\emph {flux} of photons by searching for photon candidates and
relating their number to the well-known exposure of the surface array.
This avoids the need of simulating events inititated by nuclear primaries;
only the photon energy scale is needed which can be simulated with
much higher confidence.
Two observables of the surface detectors are chosen which have
significantly different
behavior for nuclear primaries when compared to photons:
the risetime of the recorded shower signal and the radius of
curvature of the shower front.

We also derive a limit to the fraction of photons.
While the challenge of using two energy scales remains for this part
of the analysis, hadron simulations can still be avoided by using
the hybrid calibration~\cite{Roth_2007in} to reconstruct the energies
of the observed events.

The plan of the paper is as follows. In Section~\ref{sec_go}, the
observables used in the analysis and their relationship with the composition of
cosmic rays are explained. In Section~\ref{sec_photon}, the simulation of
UHE photons is considered. The method developed to
distinguish events which are photon candidates using observables of
the surface detector is detailed in Section~\ref{sec_method}. 
In Section~\ref{sec_results}, the results are presented. 
The conclusions are given in Section~\ref{sec_conclusions}.

%%%%%%%%%%%%%%%%%%%%%%%
% SECTION OBSERVABLES %
\section{Observables\label{sec_go}}
The analysis in this paper is based on data taken during 21,400 hours
of operation of the surface detector recorded in the period
1 January 2004 to 31 December 2006.
The surface detector, when completed, will
have 1600 water Cherenkov detectors spaced 1.5~km apart and covering
$\sim$3000~km$^2$~\cite{auger,sd_2007}. Each water Cherenkov detector, or station,
is a cylinder 1.2~m in height and 3.6~m in diameter. Each detector is lined
with a reflective container that holds 12~tonnes of purified water and is
fitted with three nine-inch photomultiplier tubes (PMTs) looking down into
the water.

When a relativistic particle passes through a station, Cherenkov radiation is
emitted. The radiated photons then propagate through the water, being reflected
at the station walls, and are either eventually absorbed or detected by a PMT.
The signals from the PMTs are digitised by a flash analog to digital converter
(FADC) which samples the signal every 25~ns. These digitised
signals are then transmitted to a central data acquisition system where event
triggers are built. Each event, then, has a detailed time profile
$s_i (r_i,t)$
of the energy deposited in each station $i$ at distance $r_i$ in the shower
plane.
The function $s (r,t)$ depends in a complex way both on
the parameters of the primary particle (energy, type, direction) and on
the detector response to different secondary particles (particularly the
electromagnetic and muonic shower components).

In this work, we extract two relatively simple but robust observables
from these data, noting that the wealth of information contained in
the time profiles can further be exploited in future work.
The observables, the radius of curvature of the shower front and the
risetime at 1000~m core distance, were found to provide good discrimination
between photon and nuclear primaries (see e.g.~also 
Ref.~\cite{bertou_photon}).
In addition to the quantitative studies of these observables by means of
the simulation-reconstruction chain, we will also sketch (in a simplified way)
why these observables are indeed expected to differ between nuclear and
photon primaries.

\subsection{Radius of Curvature\label{sec_curv}}

\begin{figure}[t]
  \setlength{\unitlength}{0.01\linewidth}
    \epsfig{file=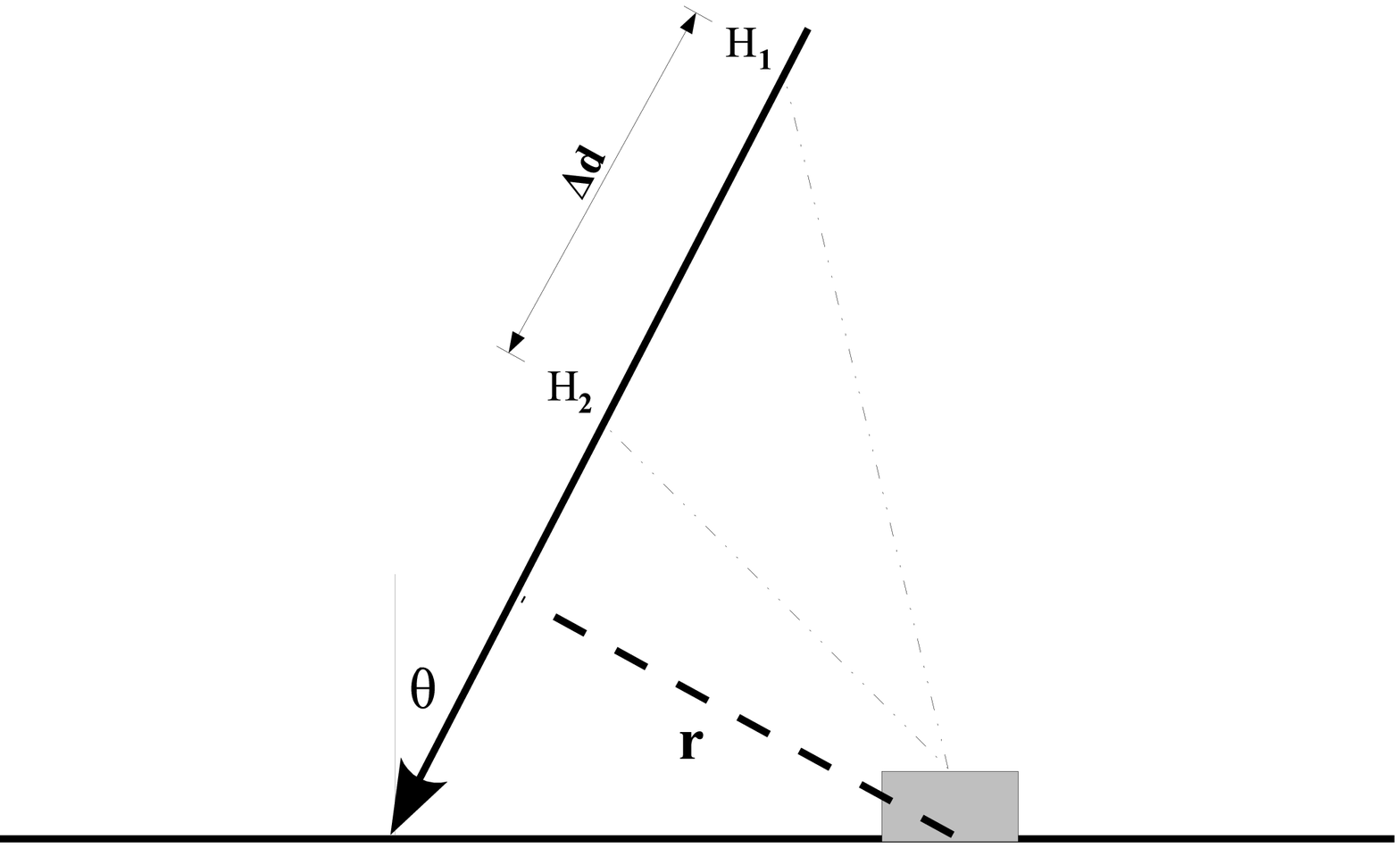,width=50\unitlength}\hfill
    \epsfig{file=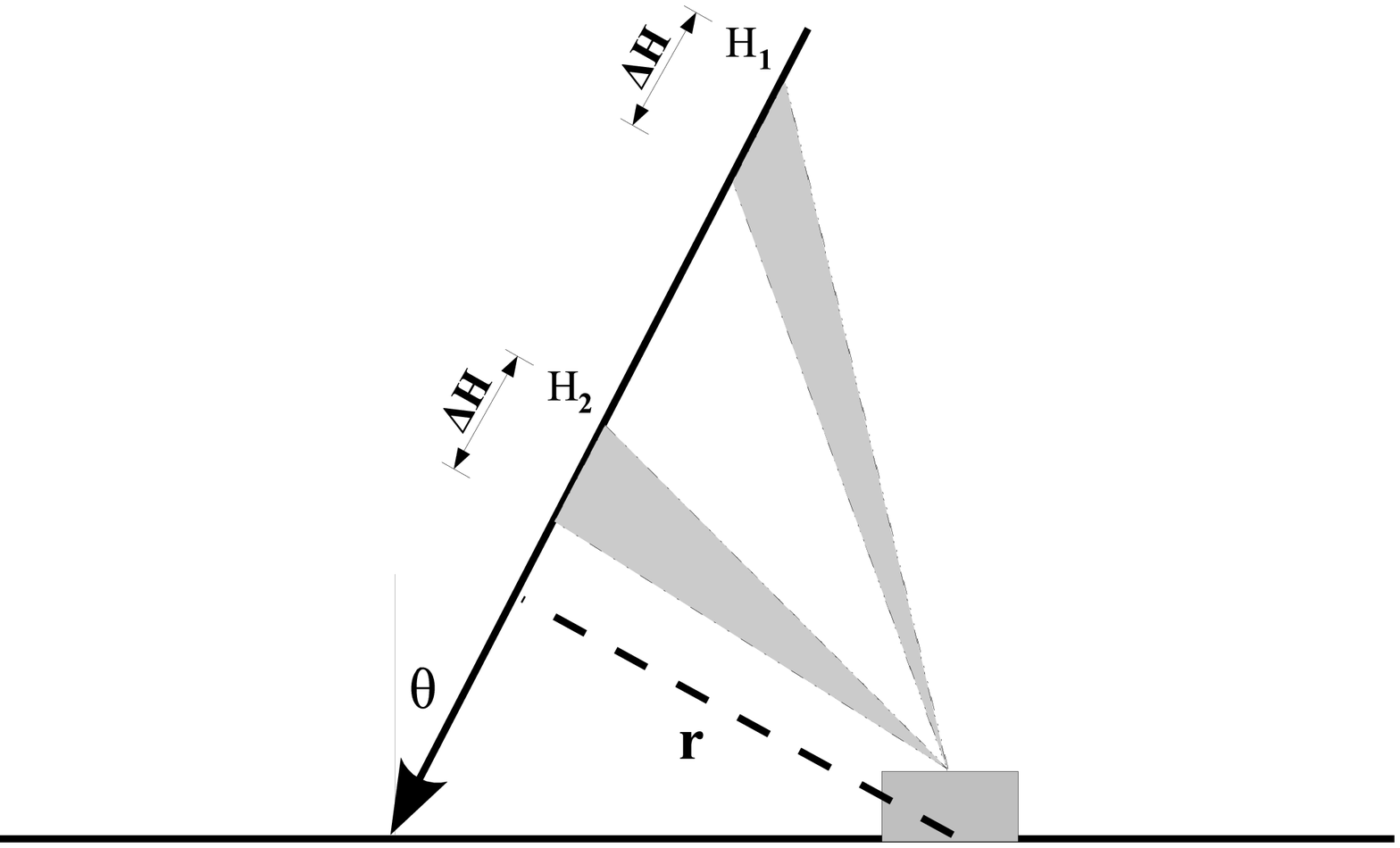,width=50\unitlength}
  \caption[risetime vs. core distance]{Illustration of geometrical effects
    on radius of curvature and risetime of the shower front.
    (Left) With respect to an imaginary planar shower front, particles
    arrive more delayed at distance $r$ when originating from a smaller
    height $H_2 < H_1$. Correspondingly, the radius of curvature of the
    actual shower front is smaller in case of the deep developing
    photon primaries.
    (Right) The spread of arrival times of particles produced over a
    pathlength $\Delta H$ and arriving at distance $r$ increases 
    for a smaller production height $H_2 < H_1$.
    Correspondingly, the risetime of the shower is increased 
    in case of the deep developing photon primaries.
    \label{rise_distance}}
\end{figure}

Due to geometrical reasons, the arrival of the first particles at
lateral distance $r$ from the axis is expected
to be delayed with respect to an (imaginary) planar shower front
(see also Fig.~\ref{rise_distance}, left plot).
For a particle that is due to an earlier interaction at
height $H$ along the shower axis and observed at $r$,
the delay from the longer path length can be approximated as
\begin{equation}
 t = \frac{1}{c} (\sqrt{H^2 + r^2} - H)~
 \propto \frac{r^2}{H} ~~ (r \ll H).
\label{t_eqn}
\end{equation}

The delay increases (for $r \ll H$ about quadratically) with  $r$.
Importantly, the delay decreases with increasing height $H$.
Air showers with the first ground particles coming from relatively
large heights will have smaller delays $t$ at fixed distance $r$
compared to showers where the registered particles originated
from smaller heights.
Compared to primary photons, showers from nuclear primaries develop
higher in the atmosphere (smaller $X_{\rm max}$).
Additionally, shower muons (much more abundant in showers from
nuclear primaries) can reach the ground from still higher altitudes
further reducing the time delay.
Thus, for nuclear primaries smaller delays are expected compared
to photon primaries.

We make use of this relation by fitting a shower front (abstract
surface with convex curvature defined by the fastest shower particles)
to the measured trigger times $t_i(r_i)$ of the first particles
registered at distances $r_i$.
In the present study, the shape of the shower front is approximated
using a spherical model (in accord with Eq.~(\ref{t_eqn})),
and the radius of curvature $R$ of the
shower front is obtained by minimizing $\chi^2$ in the function
\begin{equation}
\chi^2 = \sum_{i}\frac{[c(t_i-t_0)-|R\vec{a}-\vec{x}_i|]^2}{c^2\sigma_t^2}
\label{curv_eqn}
\end{equation}
where $t_i$ is the trigger time for station $i$ as defined
in~\cite{sdcal}, $t_0$ is the time of the shower in the center of
curvature, $\vec{a}$ is the unit vector along the shower axis,
$\vec{x}_i$ is the location of the station on the ground relative to
the shower core, and $\sigma_t$ is the uncertainty in the shower
arrival time~\cite{icrc_angle}.
In the determination of $t_i$, a software filter is applied
to reduce contributions from spurious signals not
related to the actual shower.

\subsection{Risetime}

Also the \emph{spread} in time of the signal $s_i (r_i,t)$
registered at distance $r_i$, which corresponds to the thickness of the
local shower disk, can be extracted. 
Using Eq.~(\ref{t_eqn}), the difference of arrival times of particles
originating from a height interval [$H_1, H_1 - \Delta H$] follows as

\begin{eqnarray}
\Delta t(H_1, \Delta H) 
 & \propto & r^2 \left(\frac{1}{H_1- \Delta H}-\frac{1}{H_1} \right)
  = \frac{r^2 \Delta H}{H_1(H_1-\Delta H)}
 \nonumber \\ \nonumber \\
   &<&~ \Delta t(H_2, \Delta H)
  ~~~\mbox{for}~ H_2 < H_1.
\label{dt_eqn}
\end{eqnarray}

The spread of arrival times of these particles at fixed core distance
increases for smaller production heights 
(see also Fig.~\ref{rise_distance}, right plot).
Accordingly, a larger spread is expected in case of the deep developing
photon primaries (larger $X_{\rm max}$).
We note that in general, the situation is more complex. The time spread
may depend on details of the previous shower development, particularly
also on the competition between the signals from the electromagnetic
and muonic shower components which will be commented on below.
Still, geometrical effects
are essential in the relation between time spread and primary composition.

In this study, we use the risetime $t_{1/2}(1000)$ of the shower signal
reconstructed for 1000~m distance  and located along the line given by
the projection of the shower axis onto the ground.
First, the risetime $t_{1/2}^{\rm meas}(r_i)$ of a single station is
defined as the time it takes to increase from 10\%
to 50\% of the total signal deposited in that station.
According to Eq.~(\ref{dt_eqn}), for non-vertical showers
a (moderate) dependence 
of $t_{1/2}^{\rm meas}(r_i)$ on the internal azimuth angle of the
stations within the shower plane is expected. This is because the height $H$
measured along the shower axis is larger for those stations on the exterior
side of the shower compared to those on the interior side of the shower.
To account for this, the observed $t_{1/2}^{\rm meas}(r_i)$ are
corrected depending on the internal azimuth angle $\zeta$ of that station:
\begin{eqnarray}
t_{1/2}^{\rm cor}(r_i) & = & t_{1/2}^{\rm meas}(r_i) - g\cdot\cos\zeta \label{rt_asym_eqn} \\
g & = & -66.61 + 95.13\cdot\sec\theta - 30.73\cdot\sec^2\theta + [0.001993\cdot\sec\theta \nonumber \\
 & & - 0.001259\cdot\sec^2\theta + 0.0002546\cdot\sec^3\theta - 0.0009721]\cdot r_i^2 \nonumber
\end{eqnarray}
where the parameter $g$ depends on distance $r$ and primary zenith
angle $\theta$ and is parameterised from the data,
and $\zeta$ is the clockwise angle between the projection of
the shower axis on the ground and the line connecting the shower
impact point and the station.

It is also expected from Eq.~(\ref{dt_eqn}) that the
values $t_{1/2}^{\rm cor}(r_i)$ depend on the distance $r_i$ of the stations.
We obtain the final risetime $t_{1/2}(1000)$ of the shower
by performing a fit to $t_{1/2}^{\rm cor}(r_i)$ using the function
\begin{equation}
t_{1/2}(r) = (40 + ar + br^2)\hspace{0.8ex}\mbox{ns}~.
\label{rt_eqn}
\end{equation}
The parameters $a$ and $b$ are determined for each event by fitting the
station data (typical values are 50~ns~km$^{-1}$ and
100~ns~km$^{-2}$ respectively). The function is anchored at 40~ns
at $r$=0 as that is the mean single particle response in the water
Cherenkov detectors.

While geometrical effects connected to the different shower developments
from nuclear and photon primaries are a main reason for
the risetime difference (larger $t_{1/2}(1000)$ in photon showers),
again this sensitivity to composition is further strengthened 
by shower mouns which are more abundant in the case of nuclear primaries
and can dominate the registered signal at larger zenith angles.
As muons tend to arrive within a shorter time window
compared to the electromagnetic component which suffers from multiple
scattering, this further reduces the risetime $t_{1/2}(1000)$ for
nuclear primaries.

\subsection{Energy\label{sec_energy}}

As an energy estimator, the time-integrated energy deposit $S(1000)$
at 1000~m core distance is used~\cite{s1000}.
However, for the same initial energy and direction the average $S(1000)$
from primary photons can be a factor $\ge$2 below that from nuclear
primaries~\cite{sommers,photon_energy}.
Reasons are the (typically factor $\sim$4) smaller number of muons
and, due to the later development, the steeper ground lateral distribution
in primary photon showers.
For a limit to the fraction of primary photons,
the energy scales (transformation from $S(1000)$ to primary energy) for
both photon and nuclear primaries are required, while the determination
of a limit to the flux can rely on the photon energy scale alone.

The energy scale for nuclear primaries is based on the fluorescence
technique by using events that are detected with both the surface detector
and the fluorescence telescopes~\cite{icrc_hybrid}.
The energy scale for photon primaries (which induce almost purely
electromagnetic cascades) is taken from simulations.
Thus, both approaches are largely independent from assumptions about
hadron interactions at high energy.

Using a direct relationship between $S(1000)$  and primary energy
for the photon energy scale results in a (relatively poor)
resolution of about 40\%.
To improve this, a unique energy conversion for photons is applied
that is described in detail in Ref.~\cite{photon_energy}.
It is based on the universality of shower development~\cite{icrc_uni},
i.e.~the electromagnetic part of the shower is expected to develop
in a well-predictable manner for depths exceeding $X_{\rm max}$.
In brief, for given values of $S(1000)$ and $X_{\rm max}$, 
the primary energy is estimated by
\begin{equation}
  \frac{S(1000)}{E_{\gamma}} = 1.4(1+\frac{\Delta X - 100}{1000})[1+(\frac{\Delta X - 100}{340})^2]^{-1}
  \label{ephoton_eqn}
\end{equation}
\vspace{-.5cm}
\begin{displaymath}
  \mbox{with}\qquad\Delta X = X_{\rm ground} - X_{\rm max}~,
\end{displaymath}
where $S(1000)$ is measured in units of vertical equivalent muons
(VEM)~\cite{sdcal}, the photon energy $E_{\gamma}$ is in EeV,
and $\Delta X$ is in g~cm$^{-2}$.
Since $X_{\rm max}$ is not directly measured by the surface detector alone,
an iterative approach using Eq.~(\ref{ephoton_eqn}) is taken to 
estimate the energy.
After an initial guess of the photon energy using $S(1000)$ alone,
the typical $X_{\rm max}$ of the photon showers at this energy is
taken from simulations.
With this estimate of $X_{\rm max}$, a new estimate of the photon energy
is obtained using Eq.~(\ref{ephoton_eqn}), and the procedure is repeated.
The energy estimate is found stable after few iterations and an
energy resolution of $\sim$25\% is achieved~\cite{photon_energy}.
We use this improved estimation of the photon energy, but note that
the main conclusions remain valid also when using a direct energy
estimation.
% \begin{equation}
% X_{\rm max} = (856 + 141\log E_{\gamma})\hspace{0.8ex}\mbox{g cm}^{-2}
% \label{xmax_eqn}
% \end{equation}

%%%%%%%%%%%%%%%%%%%%%%%%%%%%%%
% SECTION PHOTON SIMULATIONS %
\section{Monte Carlo Simulations\label{sec_photon}}

The QED processes of LPM effect~\cite{lpm} and geomagnetic cascading
(\cite{preshower,bertou_photon} and references therein)
need to be considered for photon showers at highest energy.
As mentioned before, the LPM effect leads to a suppression of
the pair production and bremsstrahlung cross-sections and, thus,
additionally increases the separation of photon and nuclear primaries
in terms of $X_{\rm max}$ (for a review of the LPM effect and
experimental observations of the LPM suppression,
see~\cite{klein}).\footnote{Even when artificially switching off the LPM
effect, photon showers still have a significantly larger $X_{\rm max}$
than nuclear primaries (differences $>$150~g~cm$^2$ above
10$^{19}$~eV) and a smaller number of muons.}
In case of geomagnetic cascading of UHE photons,
the initial conversion of the UHE photon into an
electron-positron pair can induce a ``preshower''
(mostly synchrotron photons plus electron-posi\-tron pair(s))
outside the atmosphere.
The subsequent air showers from such ``converted'' photons
develop higher in the atmosphere (smaller $X_{\rm max}$)
than air showers directly initiated by UHE photons do.
As geomagnetic cascading becomes important at energies above $\sim$50~EeV
at the southern site of the Auger Observatory, this process is
of minor relevance for the bulk of data used in this analysis.

The shower simulations were generated with the Aires
simulation package (v2.8), which includes the LPM effect and
geomagnetic cascading~\cite{aires}. QGS\-JET~01~\cite{qgsjet} was used as the
hadronic interaction model.
The simulation of the water Cherenkov detectors uses the
GEANT4~\cite{geant4} simulation package along with specific code that
handles PMT response and data acquisition 
electronics. The result is that the output of a simulated event is
in a format that is identical to the data format recorded with the Auger
Observatory. The shower reconstruction procedure used is the same for real
events as it is for simulated events to avoid systematic differences
at the reconstruction stage.

%%%%%%%%%%%%%%%%%%
% SECTION METHOD %
\section{Method\label{sec_method}}

In brief, the limit to the photon flux is obtained as follows.
Selection cuts are applied to the data (and simulations) to
ensure events of good reconstruction quality and a high
acceptance of the detector to photons. Based on $S(1000)$,
showers above a minimum primary energy are selected.
This data set is then searched for photon candidates using
$t_{1/2}(1000)$ and $R$ (see Section~\ref{sec_go} for definitions).
Simulations assuming photons
are used to determine the corresponding selection efficiencies.
From the number of photon candidates, the efficiencies with
respect to photons, and the experimental exposure (obtained
from the geometrical acceptance known from detector monitoring),
the upper limit to the photon flux is derived.

The criteria to select events of good quality are:

\begin{itemize}
\item the station with the largest signal is surrounded by
         6 active stations;
\item $\ge$5 stations used in the fitting of the lateral
         distribution function~\cite{icrc_ldf} out of which
         $\ge$4 stations have a non-saturated signal of $\ge$10~VEM
         (vertical equivalent muons)~\cite{sdcal};
\item reduced $\chi^2 < 10$ ($\chi^2$ from Eq.~(\ref{curv_eqn})).
\end{itemize}

The first cut restricts the analysis to well-contained events,
eliminating in particular events near the border of the array.
It affects the geometrical acceptance only.
The multiplicity criterion in the second cut is important also
to ensure a good reconstruction of $t_{1/2}(1000)$ and $R$.
As the multiplicity is related to primary
energy, this cut also affects the energy-dependent acceptance
of the array to photons.
The third cut rejects the extreme tail of the $\chi^2$ distribution
when reconstructing $R$, removing $\sim$4\% of data.
As noted before, the assumption of a spherical model used in
Eq.~(\ref{curv_eqn}) is a simplification and, thus,
not expected to provide a perfect description of the
complex features of the shower front.
This cut restricts the analysis to events where a single value
of $R$ can be reasonably extracted.
It has been checked with simulations that no bias to photons
is introduced this way.

% The distribution of reduced $\chi^2$ is also largely energy
% and angle independent within the fiducial region.
% The effect of releasing this cut is discussed below.

\begin{figure}[t]
  \setlength{\unitlength}{0.01\linewidth}
  \begin{picture}(100,65)
    \put(5,0){\includegraphics[width=90\unitlength]{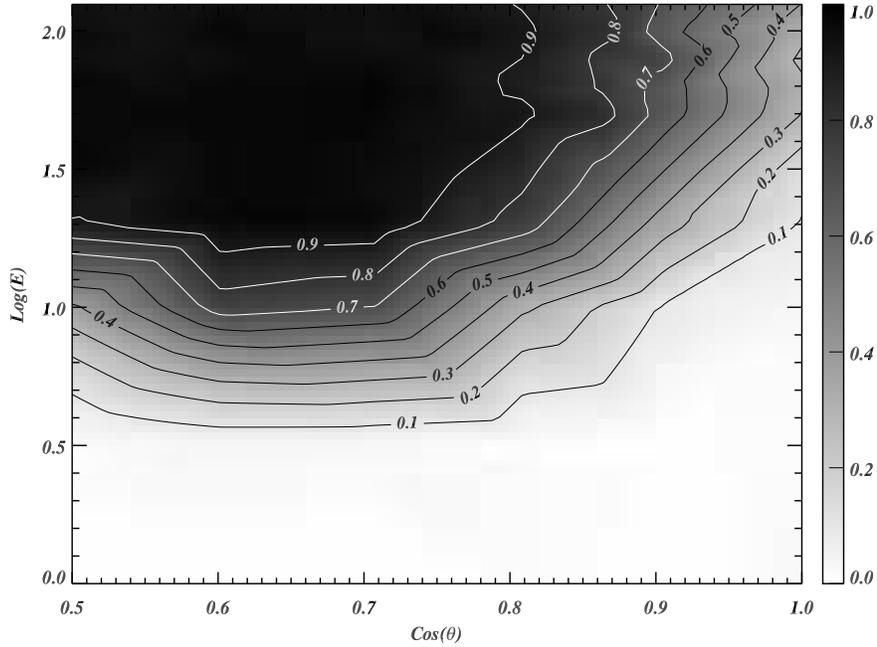}}
  \end{picture}
  \caption[Photon Detection Efficiency]{Photon detection and reconstruction
    efficiency (right hand scale) as a function of the energy (in EeV) and
    zenith angle of the primary photon. The analysis is restricted to a minimum
    energy of 10~EeV and zenith angles greater than 30$^{\circ}$ and less
    than 60$^{\circ}$ ($0.866 > \cos{\theta} > 0.5$).
    \label{efficiency}}
\end{figure}

As can be seen from Fig.~\ref{efficiency}, the resulting
photon efficiency drops to small values below $\sim$10~EeV.
At higher energy, near-vertical photons can also fail
the station multiplicity cut due to their deep development.
Therefore, the analysis is restricted to
\begin{itemize}
\item primary energies $\ge$10~EeV;
\item primary zenith angles of 30$-$60$^\circ$.
\end{itemize}

Events with zenith angles below 60$^\circ$ are selected here
since inclined showers require dedicated algorithms for an optimum
reconstruction~\cite{inclined} (this cut might be relaxed in the
future).

\begin{figure}[t]
  \setlength{\unitlength}{0.01\linewidth}
  \begin{center}
    \epsfig{file=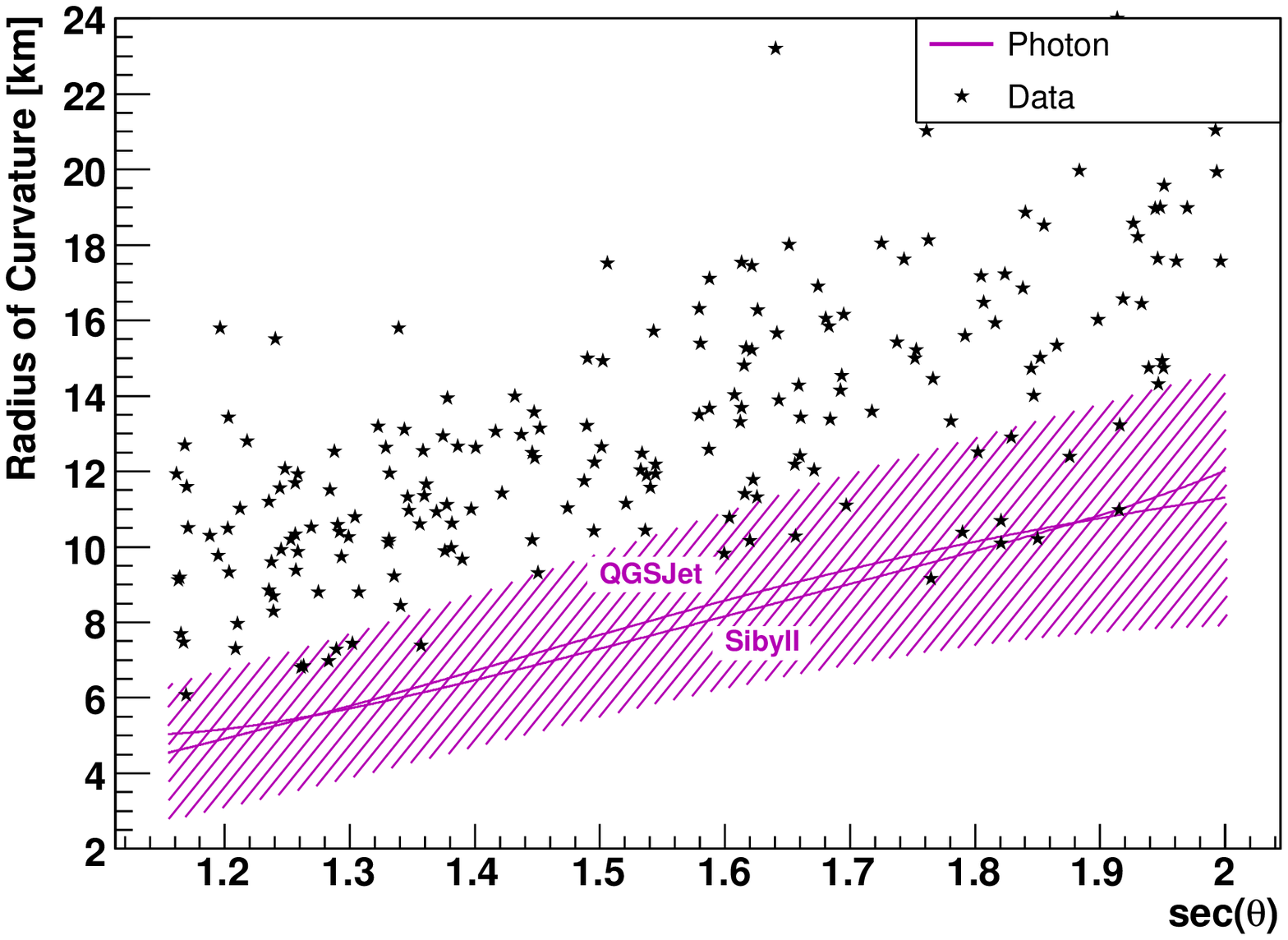,width=50\unitlength}\hfill
    \epsfig{file=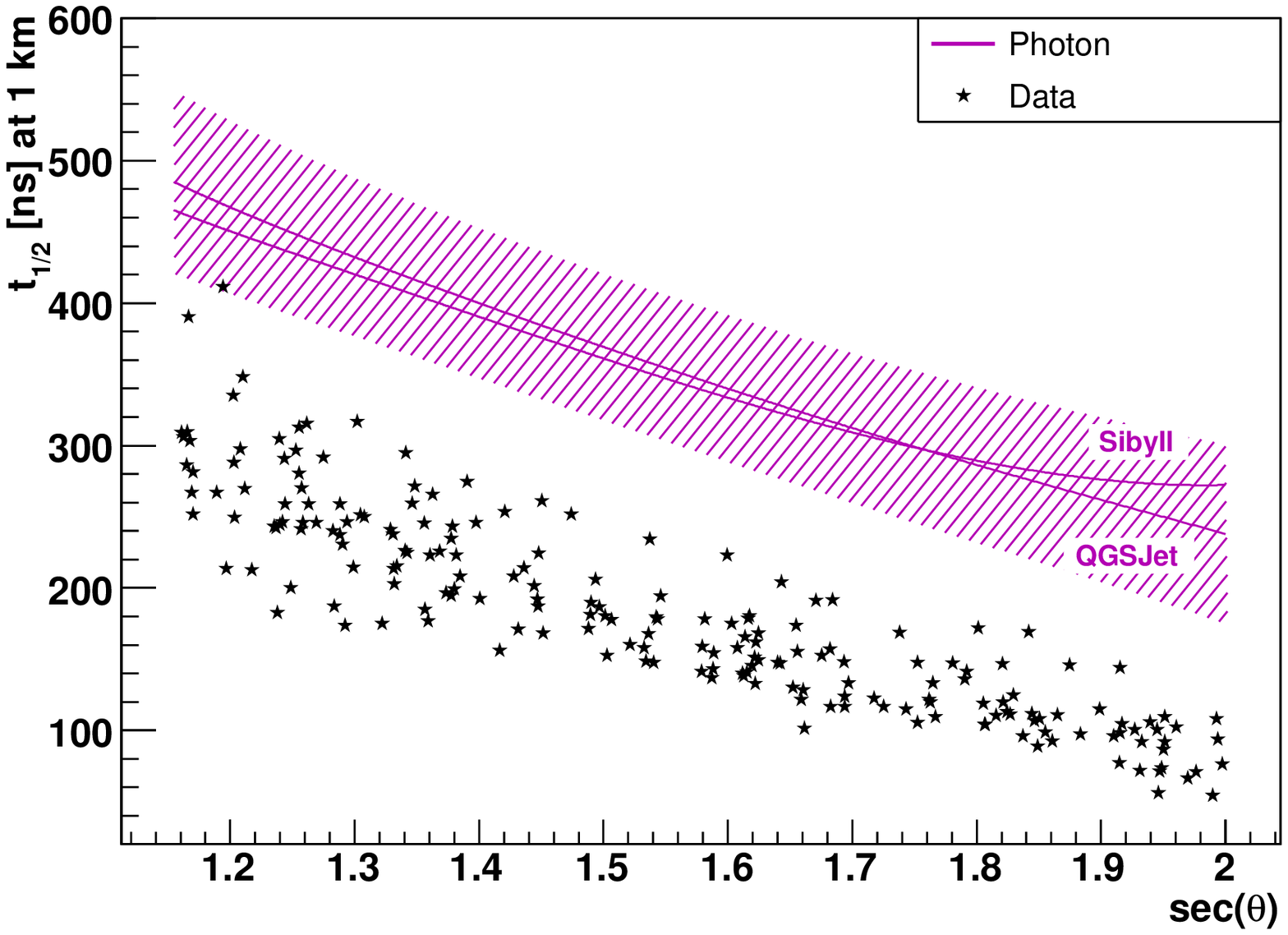,width=50\unitlength}
  \end{center}
\caption{Parameterization of the mean behavior of $R$ and $t_{1/2}$
  for 20~EeV primary photons as a function of the zenith angle
  using QGSJET~01~\cite{qgsjet} or SIBYLL~2.1~\cite{sibyll}.
  The rms values are indicated for the case of QGSJET~01.
  An increase (a decrease) of $R$ (of $t_{1/2}$) with zenith angle
  is qualitatively expected from Eqs.~(\ref{t_eqn}) and (\ref{dt_eqn})
  due to the generally longer path lengths to ground in case of
  larger inclination.
  Real events of 19--21~EeV (photon energy scale) are added.
  The significant deviation of the observed values from those expected
  for primary photons is visible.
}
\label{fig-risetime_curvature}
\end{figure}

The search for photon candidates makes use of $t_{1/2}(1000)$ and $R$ 
and consists of the following steps.
Firstly, the deviation $\Delta_x$ of the observable $x$ 
(with $x = t_{1/2}$ or $R$ referring to risetime or radius of curvature,
respectively) from the mean value $\bar{x}_{\gamma}$
predicted for photons is derived in units of the 
spread $\sigma_{x,\gamma}$ of the observable $x$,
\begin{equation}
\Delta_x =
\frac{x-\bar{x}_{\gamma}(S(1000),\theta)}{\sigma_{x,\gamma}(S(1000),\theta)}.
\label{dev_eqn}
\end{equation}
where $\bar{x}_{\gamma}(S(1000),\theta)$ and $\sigma_{x,\gamma}(S(1000),\theta)$ 
are parameterized from simulations using primary photons.
In Fig.~\ref{fig-risetime_curvature}, examples are shown for these
parameterizations of the observables along with distributions
of real events.

\begin{figure}[tbh]
  \setlength{\unitlength}{0.01\linewidth}
  \begin{picture}(100,72)
    \put(10,0){\includegraphics[width=80\unitlength]{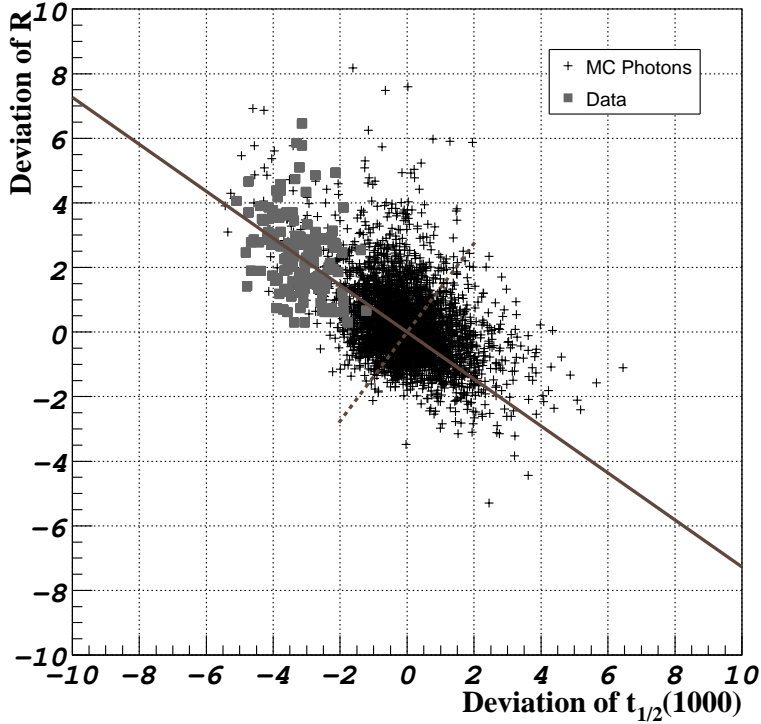}}
  \end{picture}
  \caption[Deviation from photon prediction]{The deviation from a photon
    prediction for 5\% of the data (closed squares) and simulated photon events
    (crosses). The solid grey line is the principal component axis
    identified using the limited set of real showers while the dashed
    line is the axis perpendicular to the principal component. The
    minimum energy is 10~EeV ($E_{\gamma}>10$~EeV).
    \label{2dpca}}
\end{figure}

Secondly, we combine the information contained in the quantities
$\Delta_{t_{1/2}}$ and $\Delta_R$
by performing a principal component analysis~\cite{pca_oja},
leaving a more sophisticated statistical analysis for the future.
To determine the principal component (defined as the axis with
the largest variance),  5\% of the real events are used together
with results from photon simulations, see Fig.~\ref{2dpca}.
For the simulations, a power law spectrum of index -2.0 has been
assumed (see below for other indices).
The remaining 95\% of the data are then projected onto the principal
axis along with the simulated photons.

This procedure allows the \textit{a priori} definition of a simple cut
in the projected distribution to finally obtain photon candidate events.
The cut was chosen at the mean of the distribution for photons,
such that the efficiency of this cut is $f=0.5$ by construction.
Any real event falling above this cut will be considered a photon
candidate.
We note that such photon candidates, if occuring, can not
yet be considered as being photons, as they actually might be due
to background events from nuclear primaries.
A presence of background events would result in weaker upper limits
(larger numerical values) in the analysis approach adopted here.
% We do not
%subtract any such background in the present analysis to (again) avoid
%relying on simulations of hadron interaction.

Finally,
an upper limit on the number of photons
$\mathcal{N}_{\gamma}^{\rm CL}$ at confidence level CL 
is calculated from the number
of photon candidate events $N_{\gamma}$ above a minimum energy, $E_{\rm min}$.
The upper limit on the flux or fraction of photons above a given energy is
based on $\mathcal{N}_{\gamma}^{\rm CL}$ along with the integrated efficiency
$\varepsilon$ of accepting photons,
the photon selection cut
efficiency  ($f=0.5$), and either the exposure $A$ of the detector for
the flux limit:
\begin{equation}
\Phi_{\rm CL}(E>E_{\rm min})
 = \frac{\mathcal{N}_{\gamma}^{\rm CL}(E_{\gamma}>E_{\rm min})
\times\frac{1}{f}\times\frac{1}{\varepsilon}}{0.95A}~,
\label{flux_eqn}
\end{equation}
or the number of non-photon candidate events $N_{\rm non-\gamma}$ in the data set
for the fraction limit:
\begin{equation}
\mathcal{F}_{\rm CL}(E>E_{\rm min}) = \frac{\mathcal{N}_{\gamma}^{\rm CL}(E_{\gamma}>E_{\rm min})
\times\frac{1}{f}\times\frac{1}{\varepsilon}}
{N_{\gamma}(E_{\gamma}>E_{\rm min})+N_{\rm non-\gamma}(E_{\rm non-\gamma}>E_{\rm min})}~.
\label{frac_eqn}
\end{equation}

In Eq.~(\ref{flux_eqn}), the factor 0.95 is from the fact that only 95\% of
the data are used to determine the number of photon candidate events.
The energy is labeled as either
the energy according to the photon energy reconstruction, $E_{\gamma}$, or
(required in Eq.~(\ref{frac_eqn}))
the energy according to the non-photon energy reconstruction,
$E_{\rm non-\gamma}$.

Experimentally, the limit $\Phi_{\rm CL}$ to the flux is more robust than
the limit $\mathcal{F}_{\rm CL}$ to the fraction due to the
different denominators of Eqs.~(\ref{flux_eqn}) and (\ref{frac_eqn}).
For $\mathcal{F}_{\rm CL}$, two energy scales are required; also, with
increasing energy, the statistical uncertainty of the quantity
$(N_{\gamma}+N_{\rm non-\gamma})$
becomes large. For $\Phi_{\rm CL}$, in contrast,
the aperture is known to good ($\sim$3\%) accuracy.

\begin{figure}[t]
  \setlength{\unitlength}{0.01\linewidth}
  \begin{picture}(100,72)
    \put(10,0){\includegraphics[width=80\unitlength]{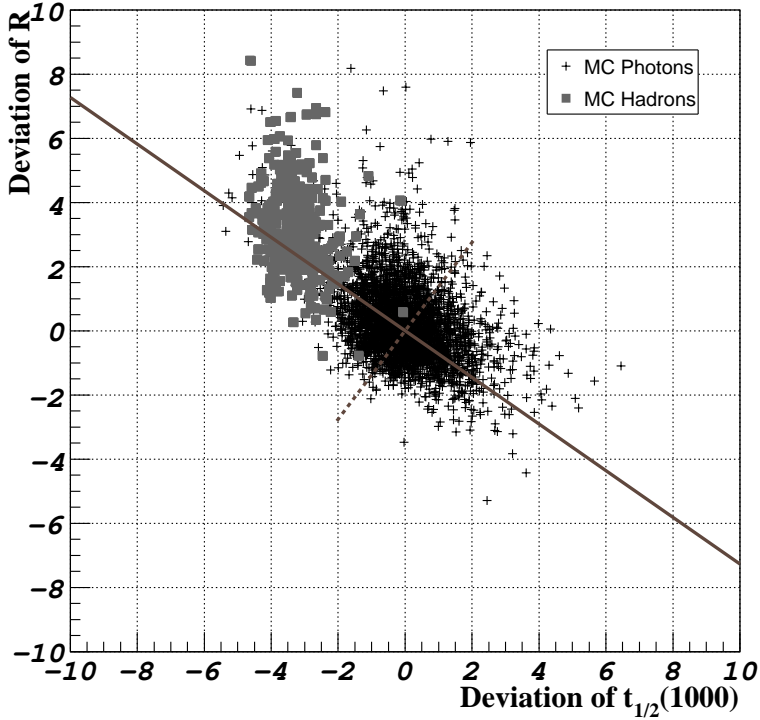}}
  \end{picture}
  \caption[Monte Carlo hadrons' deviation from photon prediction]{The
  black crosses are simulated photon showers while the squares are a
  mixture of Monte Carlo proton and iron with an energy of 10~EeV.
  For comparison, the lines shown in Fig.~\ref{2dpca} (principal
  component axes) are added.
  The distribution of simulated nuclear primaries is similar to the
  distribution of real data seen in Fig.~\ref{2dpca}.\label{2dpcaMC}}
\end{figure}

Though the present work does not aim at extracting a composition of
nuclear primaries, it is interesting to check whether the principal
component axis found
from real data and the separation along it reflects what would be
expected if the bulk of the real data is due to nuclear primaries.
In Fig.~\ref{2dpcaMC}, the same simulated photon events are used
as in Fig.~\ref{2dpca} but
the 5\% of real data are replaced with a set of $\sim$750 Monte
Carlo proton and iron showers with an energy of 10~EeV. The separation
observed in real data is both in the same direction and of a similar
magnitude as that expected from simulated nuclear primaries. 

%%%%%%%%%%%%%%%%%%%
% SECTION RESULTS %
\section{Results\label{sec_results}}

\begin{figure}[t]
  \setlength{\unitlength}{0.01\linewidth}
  \begin{picture}(100,72)
    \put(10,0){\includegraphics[width=80\unitlength]{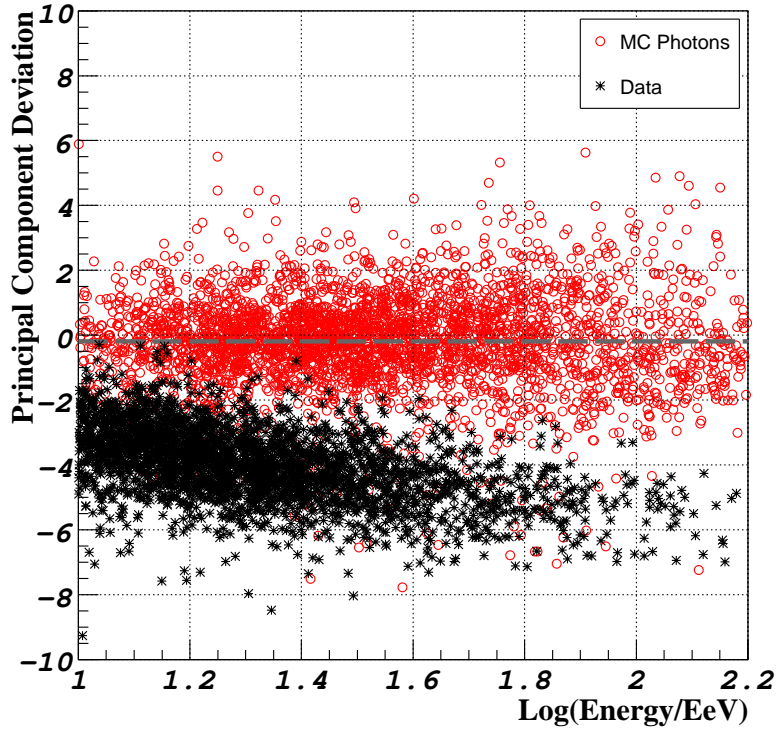}}
  \end{picture}
  \caption[Data and photon pca prediction]{The deviation of data
  (black crosses) and photons (open red circles) from the principal
  component as a function of the primary energy (photon energy scale).
  Data lying above the dashed line, which indicates the mean of the
  distribution for photons, are taken as photon candidates.
  No event meets this requirement. \label{pcascatt}}
\end{figure}

\begin{figure}
  \setlength{\unitlength}{0.01\linewidth}
  \begin{picture}(100,91)
    \put(0,47){\includegraphics[width=50\unitlength]{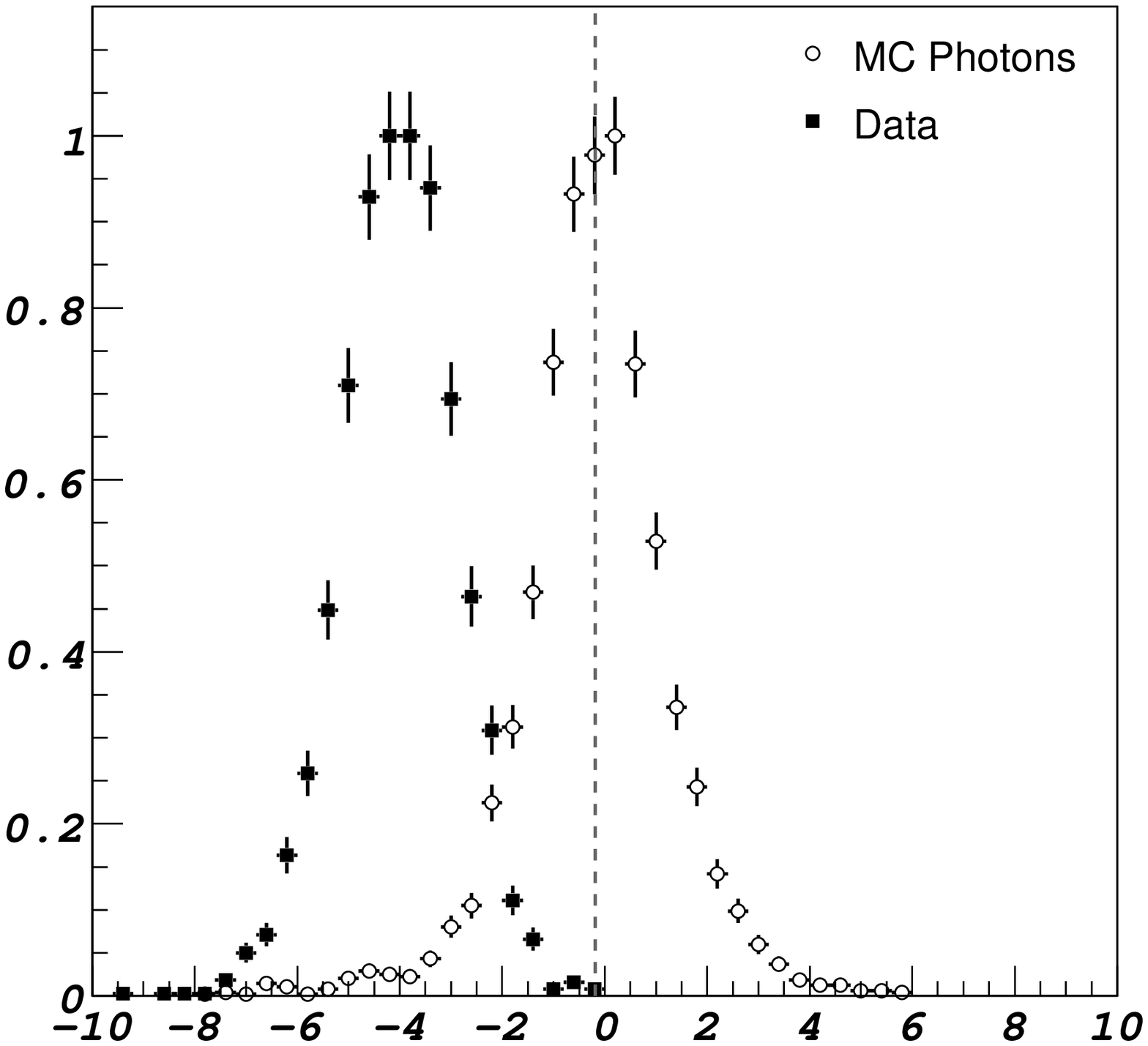}}
    \put(50,47){\includegraphics[width=50\unitlength]{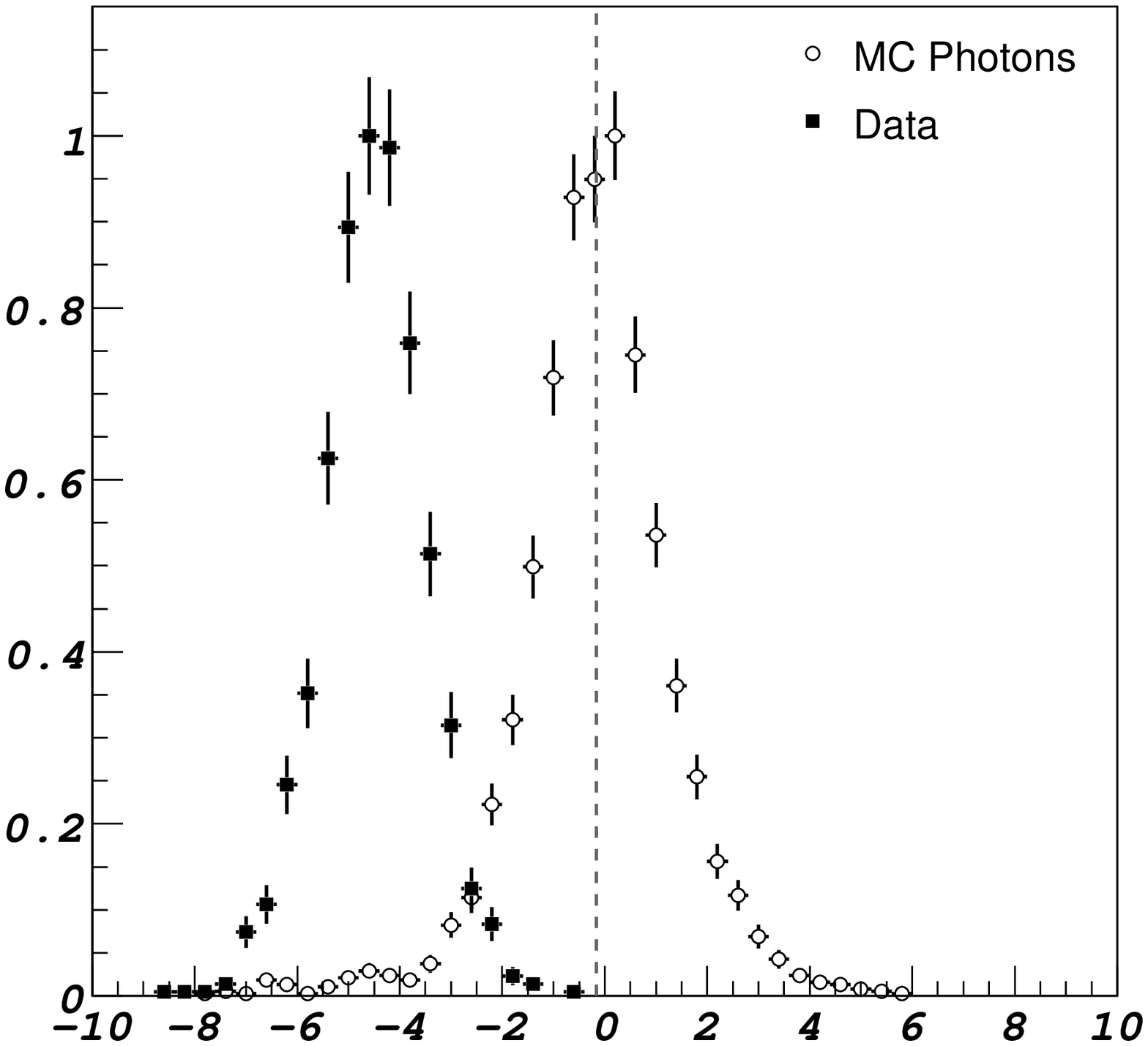}}
    \put(25,0){\includegraphics[width=50\unitlength]{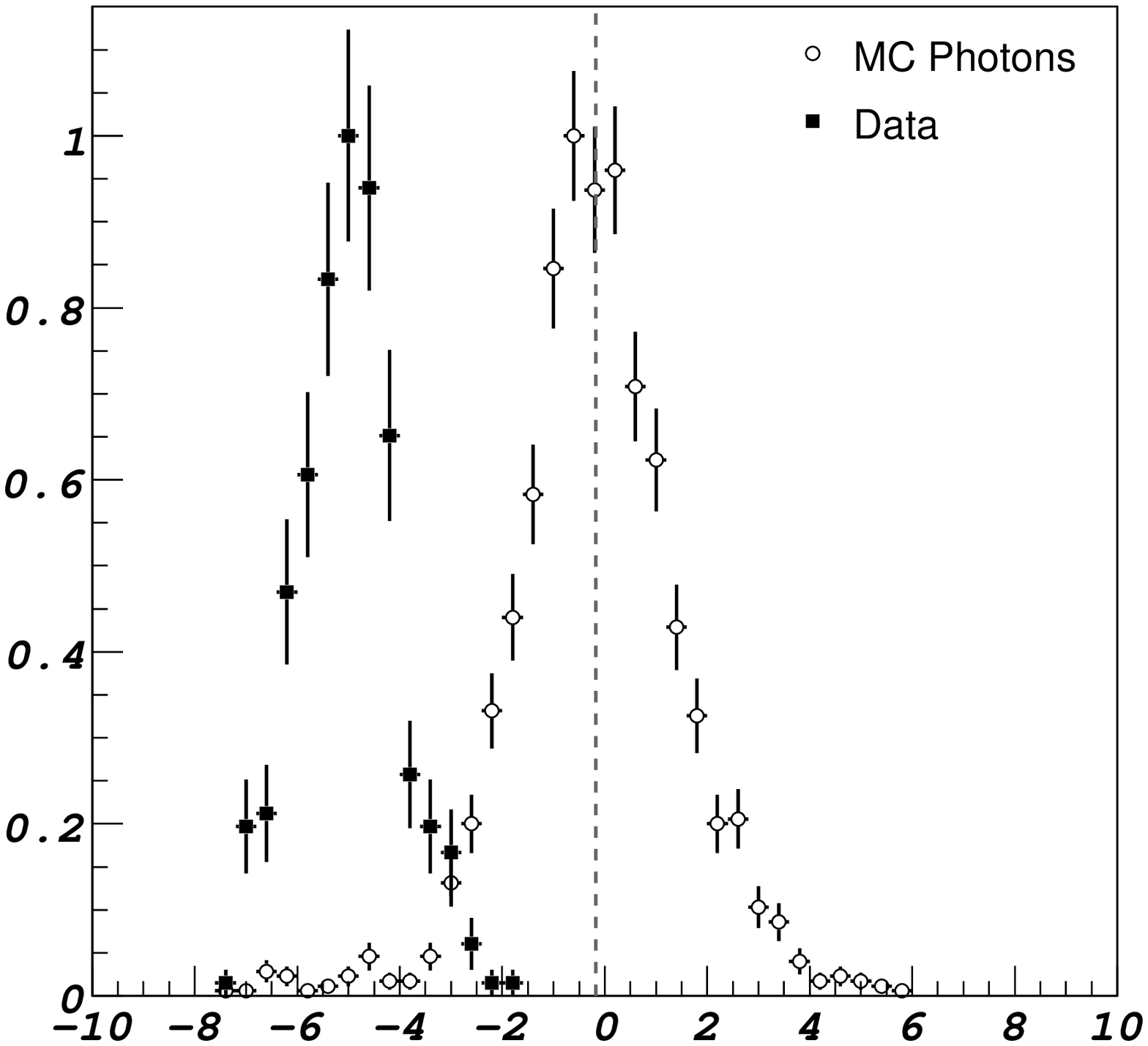}}
  \end{picture}
  \caption[Photon Search]{Distribution of real events (closed squares) along
    with simulated photon events (open circles) for the projection on the principal
    component axis. The photon candidate cut is set at the mean of the distribution
    for photons and is shown as the dotted line. The plots are made
    requiring a minimum energy (according to the photon energy converter) of
    10~EeV (top-left), 20~EeV (top-right), and 40~EeV (bottom).
    Distributions are normalised to unity at maximum.
    \label{photon}}
\end{figure}

The data from 2004--2006 are analysed as described
in the preceding section.
The integrated aperture of the Observatory 
is 3130~km$^2$~sr~yr for the angular coverage regarded in
this analysis.
Above 10, 20, and 40~EeV, for the energy scale of photons
(in brackets for nuclear primaries), the data set consists of
2761 (570), 1329 (145), and 372 (21) events.
The measured values of $t_{1/2}(1000)$ and $R$ are used to determine
the projection on the principal axis.
A scatter plot of this quantity vs.~the primary energy is shown in
Fig.~\ref{pcascatt}, while in Fig.~\ref{photon} the corresponding
distributions are plotted for the three threshold energies.
No event passes the photon candidate cut.
The upper limits on the photon flux above 10, 20, and
40~EeV are then \fluxa{}, \fluxb{}, and \fluxc{}~\fluxunit{}
(at 95\% CL).
The limits on the photon fraction are
 \fraca{}, \fracb{}, and \fracc{} (at 95\% CL)
above 10, 20, and 40~EeV.
In Tab.~\ref{table_results}, all relevant quantities
(number of events, efficiencies, resulting limits) are summarized.
%As a simple cross-check, we note that dividing $\Phi_{\rm CL}$
%by the (integral) total flux $\Phi_{\rm tot}$
%using the Auger spectrum~\cite{Roth_2007in}
%(justified for $\Phi_{\rm CL} \ll \Phi_{\rm tot}$
%gives an estimate of the limit to the photon fraction
%which agrees reasonably well to $\mathcal{F}_{\rm CL}$.

\begin{table}
  \begin{center}
    \begin{tabular}{|c|c|c|c|c|c|c|c|}
      \hline
      $E_{\rm min}$ & $N(E_{\gamma}>E_{\rm min})$ & $N_{\gamma}$ & $\mathcal{N}^{0.95}_{\gamma}$
      & $N_{\rm non-\gamma}$ &
      $\varepsilon$ & $\Phi_{0.95}$ & $\mathcal{F}_{0.95}$ \\
      \hline
      10 &\Ntesta{}&\Nphoa{}&\Lima{}&\Nnona{}&\effa{}&\fluxa{}&\fraca{} \\
      20 &\Ntestb{}&\Nphob{}&\Limb{}&\Nnonb{}&\effb{}&\fluxb{}&\fracb{} \\
      40 &\Ntestc{}&\Nphoc{}&\Limc{}&\Nnonc{}&\effc{}&\fluxc{}&\fracc{} \\
      \hline
    \end{tabular}
  \end{center}
  \caption[Photon Limit Results]{Results of the analysis searching for photon
    candidate events. The fraction and flux limits are integral limits above
    $E_{\rm min}$ (EeV), $\varepsilon$ is the efficiency of detection and
    reconstruction, $\Phi_{0.95}$ is in units of \fluxunit{}, and all results
    are 95\% confidence level.}
\label{table_results}
\end{table}

From Fig.~\ref{pcascatt} it can also be seen that the separation of
data and photon primaries increases with energy.
In particular at highest energies above $E_{\rm GZK}$ for the photon
energy scale, there is no indication for photon-initiated events.
%(we note that energies of nuclear primaries have to be reduced by
%by a factor $\sim$2 or more).
Thus, the absence of photons, within the improved limits placed in this work,
shows that the method applied by the Auger Observatory to calibrate
the shower energy is not strongly biased by a photon ``contamination''.

We studied potential sources of systematic effects in the analysis.
To determine the efficiency to photons and to establish the photon 
candidate cut, a primary photon spectrum of power law index -2.0
has been used in the simulations, motivated by predictions from top-down
models in (e.g.~in Ref.~\cite{shdm}).
The effect of changing the power law index to -1.7, -2.5, and -3.0
has been investigated. The number of events which are photon candidates is
unchanged (along with the number of non-photon candidate events),
but the correction for the photon efficiency changes.
Specifically, for a steeper input spectrum (increased fraction of
lower-energy photons), the efficiency decreases.
The summary of the results can be seen in Table~\ref{powerlaw_table}.
For 10~EeV threshold energy, limits change from 
(3.8$\rightarrow$5.5)$\times$$10^{-3}$~km$^{-2}$~sr$^{-1}$~yr$^{-1}$ for the flux
and from (2.0$\rightarrow$2.9)\% for the fraction. 
The differences get smaller with increased threshold energy.

\begin{table}
  \begin{center}
    \begin{tabular}{|c|ccc|ccc|ccc|}
      \hline
      $E_{\rm min}$ & 10 & 20 & 40 & 10 & 20 & 40 & 10 & 20 & 40 \\
      \hline
      $\alpha$&\multicolumn{3}{|c|}{Efficiency ($\varepsilon$)}&\multicolumn{3}{|c|}{Flux ($\times 10^{-3}$)}&\multicolumn{3}{|c|}{Fraction [\%]} \\
      \hline
      1.7 &\effaa{}&\effba{}&\effca{}&\Sfluxaa{}&\Sfluxba{}&\Sfluxca{}&\Sfracaa{}&\Sfracba{}&\Sfracca{} \\
      2.0 &\effa{} &\effb{} &\effc{} &\Sfluxab{}&\Sfluxbb{}&\Sfluxcb{}&\Sfracab{}&\Sfracbb{}&\Sfraccb{} \\
      2.5 &\effac{}&\effbc{}&\effcc{}&\Sfluxac{}&\Sfluxbc{}&\Sfluxcc{}&\Sfracac{}&\Sfracbc{}&\Sfraccc{} \\
      3.0 &\effad{}&\effbd{}&\effcd{}&\Sfluxad{}&\Sfluxbd{}&\Sfluxcd{}&\Sfracad{}&\Sfracbd{}&\Sfraccd{} \\
      \hline
    \end{tabular}
  \end{center}
  \caption[Power Law Differences]{Results when changing the exponent ($\alpha$)
    in the power law of the simulated spectrum. The default value is 2.0.
    The efficiency of detection and reconstruction is on the left, the
    resulting limit on the fraction of photons is on the right, and the limit
    on the integrated flux is listed in the middle in units of
    \fluxunit{} (95\% CL).}
\label{powerlaw_table}
\end{table}

The photonuclear cross-section used in the simulation is based on the
Particle Data Group (PDG) extrapolation~\cite{pdg}.
For an increased cross-section, more energy would be transfered to the
hadron (and muon) component which could diminish the separation power
between data and primary photons~\cite{risse_c2cr}.
From unitarity constraints, the cross-section is not expected to
exceed the PDG extrapolation by more than $\sim$75\% at
10~EeV~\cite{rogers}; at $10^{15}$~eV, where the
difference in cross-section would have a greater impact on the shower
development, the maximum difference is $\sim$20\%. From simulations with
modified cross-sections it was verified that this leads to a negligible
variation of the average values of the discriminating variables used
in the current analysis.

The simulations have been performed with AIRES using the QGSJET hadronic
interaction model. As a cross-check, calculations with
CORSIKA~\cite{corsika} / QGS\-JET and AIRES / SIBYLL were conducted, both of
which show reasonable agreement to the AIRES / QGSJET case.
As the cascade initiated by primary photons has an almost pure
electromagnetic nature, indeed no significant effect is expected when
changing to another interaction model.
This minor dependence of the results on the details of hadron
interactions, which are largely uncertain at high energy, is an
important advantage of searches for primary photons.

\begin{figure}
  \setlength{\unitlength}{0.01\linewidth}
  \begin{picture}(100,75)
    \put(5,0){\includegraphics[width=90\unitlength]{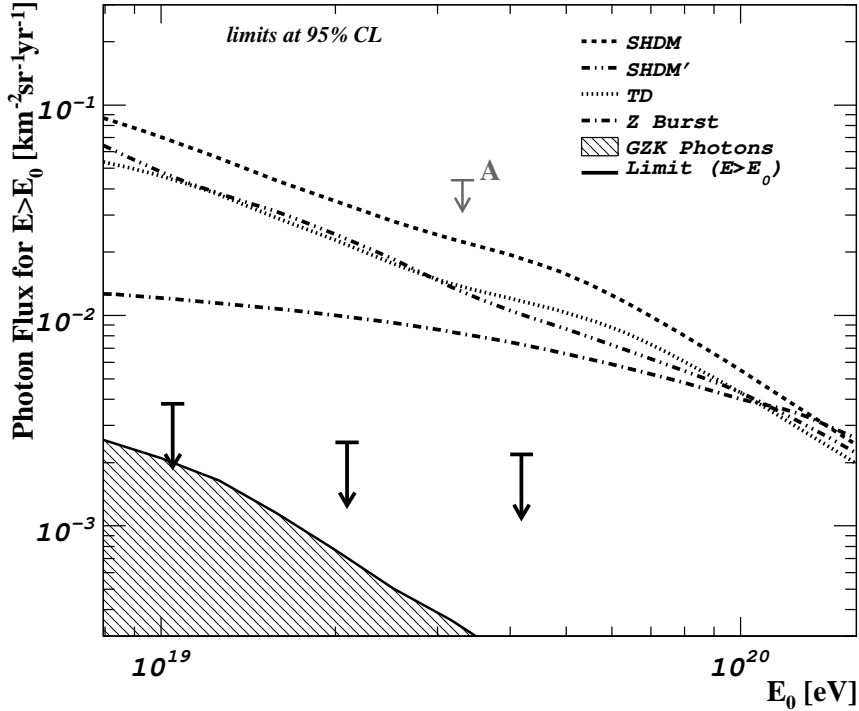}}
  \end{picture}
  \caption[Photon Flux Limit]{The upper limits on the integral flux of photons
  derived in this work (black arrows)
  along with predictions from top-down models
  (SHDM, TD and ZB from Ref.~\cite{gelmini}, 
   SHDM' from  Ref.~\cite{ellis}) and with
  predictions of the GZK photon flux~\cite{gelmini}.
    A flux limit derived indirectly by AGASA (``A'')
% (multiplying a limit to the fraction from AGASA data
% with the AGASA spectrum)
    is shown for comparison~\cite{agasa_photon}.
    \label{flux_limit}}
\end{figure}

\begin{figure}
  \setlength{\unitlength}{0.01\linewidth}
  \begin{picture}(100,75)
    \put(5,0){\includegraphics[width=90\unitlength]{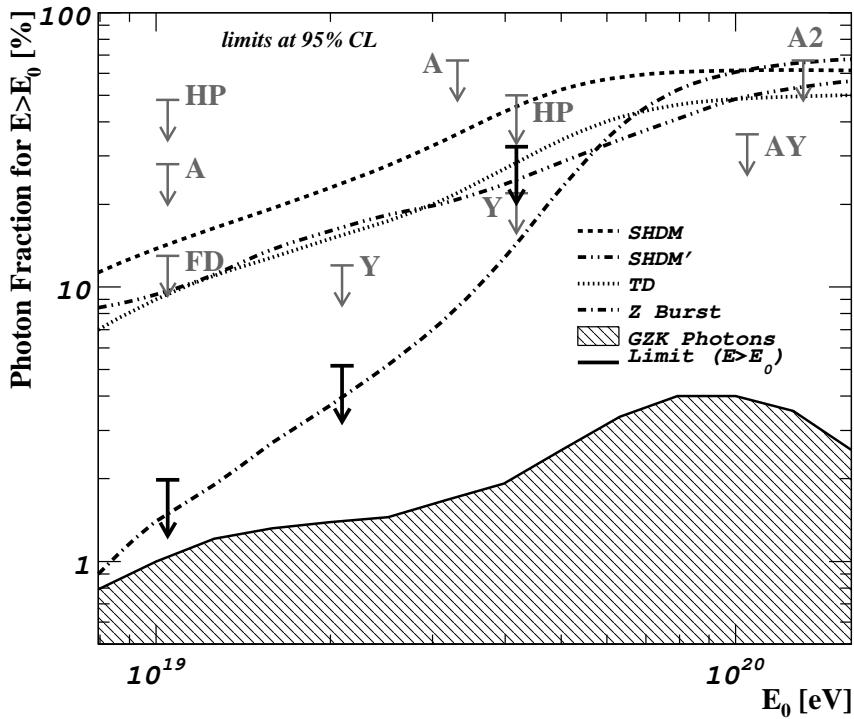}}
  \end{picture}
  \caption[Photon Fraction Limit]{The upper limits on the fraction of photons
  in the integral cosmic-ray flux
  derived in this work (black arrows)
    along with previous experimental limits 
    (HP: Haverah Park~\cite{hp_photon};
     A1, A2: AGASA~\cite{agasa_photon,risse_photon};
     AY: AGASA-Yakutsk~\cite{ay_limit};
     Y: Yakutsk~\cite{yakutsk};
     FD: Auger hybrid limit~\cite{fd_limit}).
    Also shown are predictions from top-down models
   (SHDM, TD and ZB from Ref.~\cite{gelmini},
    SHDM' from  Ref.~\cite{ellis}) 
    and predictions of the GZK photon fraction~\cite{gelmini}.
    \label{frac_limit}}
\end{figure}

The new limits are compared to previous results and to theoretical
predictions in Fig.~\ref{flux_limit} for the photon flux
and in  Fig.~\ref{frac_limit} for the photon fraction.
We placed the first direct limit to the flux of UHE photons
(an earlier bound from AGASA, about an order of magnitude weaker
than the current bounds, was derived indirectly via a limit
to the fraction and the flux spectrum~\cite{agasa_photon}).
In terms of the photon fraction, the current bound at 10~EeV
approaches the $10^{-2}$ level while previous bounds were at the
$10^{-1}$ level. 

% Main reasons for this improvements are a larger
% number of events combined with observables offering a good
% discrimination power between photons and nuclear primaries.

A discovery of a substantial photon flux could have been
interpreted as a signature of top-down models.
In turn, the experimental limits now put strong constraints on
these models.
For instance, certain SHDM or TD models discussed in the literature
(SHDM and TD from Ref.~\cite{gelmini} based on the fragmentation
calculations of  Ref.~\cite{aloisio04},
SHDM' from  Ref.~\cite{ellis}\footnote{Two others of the eight photon
flux spectra calculated in Ref.~\cite{ellis} from crypton decays may
still be compatible with our limits within a factor $\sim$2.})
predict fluxes that exceed the limits by a 
factor $\sim$10. 
It should be noted that a simple rescaling of the flux predictions
from top-down models,
which were motivated by and based on the energy spectrum observed
by AGASA,
would reduce the predicted photon flux  by only a factor $\sim$2
which would still overshoot our experimental limit by a factor $\sim$5
at $10^{19}$~eV.
While a minor contribution from top-down models to the observed
UHE cosmic-ray flux might still be allowed within the limits derived
in this work, current top-down models do not appear to provide an adequate
explanation of the origin of the highest-energy cosmic rays
(see also Ref.~\cite{semikoz_auger} for a comparison of photon flux
predictions to the Auger limits for different top-down model
parameters).

In acceleration models, photon fluxes are usually expected to be
a factor 2 or more below the current bounds (cf.~the GZK photon
predictions in the Figs.~\ref{flux_limit} and \ref{frac_limit}
from Ref.~\cite{gelmini}).
Such fluxes can be tested with future data taken at the Auger Observatory
(see also Ref.~\cite{review}).
After five years of operation with
the complete surface detector, sensitivities at the level of
$\sim$$4 \times 10^{-4}$~\fluxunit{} for the integrated flux and
$\sim$0.7\% for the fraction of photons above 20~EeV (95\% CL) could be reached.

%%%%%%%%%%%%%%%%%%%
% SECTION CONCLUSIONS %
\section{Conclusions\label{sec_conclusions}}

Using data from the surface detector we obtained 95\% c.l.~upper limits 
on the photon flux of \fluxa{}, \fluxb{}, and \fluxc{}~\fluxunit{} 
above $10^{19}$~eV, $2\times10^{19}$~eV, and $4\times10^{19}$~eV.
These are the first direct bounds on the {\it flux} of UHE photons.
For the photon fraction, limits of
\fraca{}, \fracb{}, and \fracc{} were placed.

These limit improve significantly upon bounds from previous experiments
and put strong constraints on certain models of the origin of cosmic rays.
Current top-down models such as the super-heavy dark matter scenario
do not appear to provide an adequate explanation of the UHE cosmic rays.
In bottom-up models of acceleration of nuclear primaries in
astrophysical sources, the expected photon fluxes are typically
well below the current bounds. An astrophysical origin of UHE cosmic rays
is also suggested by the recent discovery of a correlation of UHE cosmic rays
with the directions of nearby AGNs~\cite{science}.
Concerning the method of energy calibration
as applied by the Auger Observatory, the photon bounds derived in this
work show that there is no strong bias due to a contamination from UHE photons.

%Possible extensions of the analysis presented here may include, for instance,
%the usage of further information contained in the surface detector signals
%or employing more sophisticated statistical methods for separating photons
%from nuclear primaries. Related studies are ongoing.
With the data accumulating over the next years, and particularly when complementing
the Auger southern site by an extended northern one, 
the flux levels expected for GZK photons may be in reach.

%%%%%%%%%%%%%%%%%%%

\textit{Acknowledgements:}

The successful installation and commissioning of the Pierre Auger Observatory
would not have been possible without the strong commitment and effort
from the technical and administrative staff in Malarg\"ue.

We are very grateful to the following agencies and organizations for financial support: 
Comisi\'on Nacional de Energ\'ia At\'omica, Fundaci\'on Antorchas,
Gobierno De La Provincia de Mendoza, Municipalidad de Malarg\"ue,
NDM Holdings and Valle Las Le\~nas, in gratitude for their continuing
cooperation over land access, Argentina; the Australian Research Council;
Conselho Nacional de Desenvolvimento Cient\'ifico e Tecnol\'ogico (CNPq),
Financiadora de Estudos e Projetos (FINEP),
Funda\c{c}\~ao de Amparo \`a Pesquisa do Estado de Rio de Janeiro (FAPERJ),
Funda\c{c}\~ao de Amparo \`a Pesquisa do Estado de S\~ao Paulo (FAPESP),
Minist\'erio de Ci\^{e}ncia e Tecnologia (MCT), Brazil;
Ministry of Education, Youth and Sports of the Czech Republic;
Centre de Calcul IN2P3/CNRS, Centre National de la Recherche Scientifique (CNRS),
Conseil R\'egional Ile-de-France,
D\'epartement  Physique Nucl\'eaire et Corpusculaire (PNC-IN2P3/CNRS),
D\'epartement Sciences de l'Univers (SDU-INSU/CNRS), France;
Bundesministerium f\"ur Bildung und Forschung (BMBF),
Deutsche Forschungsgemeinschaft (DFG),
Finanzministerium Baden-W\"urttemberg,
Helmholtz-Gemeinschaft Deutscher Forschungszentren (HGF),
Ministerium f\"ur Wissenschaft und Forschung, Nordrhein-Westfalen,
Ministerium f\"ur Wissenschaft, Forschung und Kunst, Baden-W\"urttemberg,
Germany; Istituto Nazionale di Fisica Nucleare (INFN),
Ministero dell'Istruzione, dell'Universit\`a e della Ricerca (MIUR), Italy;
Consejo Nacional de Ciencia y Tecnolog\'ia (CONACYT), Mexico;
Ministerie van Onderwijs, Cultuur en Wetenschap,
Nederlandse Organisatie voor Wetenschappelijk Onderzoek (NWO),
Stichting voor Fundamenteel Onderzoek der Materie (FOM), Netherlands;
Ministry of Science and Higher Education,
Grant Nos. 1 P03 D 014 30, N202 090 31/0623, and PAP/218/2006, Poland;
Funda\c{c}\~ao para a Ci\^{e}ncia e a Tecnologia, Portugal;
Ministry for Higher Education, Science, and Technology,
Slovenian Research Agency, Slovenia;
Comunidad de Madrid, Consejer\'ia de Educaci\'on de la Comunidad de Castilla
La Mancha, FEDER funds, Ministerio de Educaci\'on y Ciencia,
Xunta de Galicia, Spain;
Science and Technology Facilities Council, United Kingdom;
Department of Energy, Contract No. DE-AC02-07CH11359,
National Science Foundation, Grant No. 0450696,
The Grainger Foundation USA; ALFA-EC / HELEN,
European Union 6th Framework Program,
Grant No. MEIF-CT-2005-025057, and UNESCO.


\begin{thebibliography}{99}

\bibitem{linsley}
J.~Linsley, Phys.~Rev.~Lett.~{\bf 10}, 146 (1963).
%%CITATION = PRLTA,10,146;%%

\bibitem{hp_high}
M.A.~Lawrence, R.J.O.~Reid, A.A.~Watson, J.~Phys. {\bf G17}, 733 (1991).
%%CITATION = JPHGB,G17,733;%%

\bibitem{flyseye_high}
D.J.~Bird {\it et al.}, Phys.~Rev.~Lett.~{\bf 71}, 3401 (1993).
%%CITATION = PRLTA,71,3401;%%

\bibitem{agasa_high}
N.~Hayashida {\it et al.}, Phys.~Rev.~Lett.~{\bf 73}, 3491 (1994).
%%CITATION = PRLTA,73,3491;%%

\bibitem{hires-gzk}
R.U.~Abbasi {\it et al.}, Phys.~Lett.~B {\bf 619}, 271 (2005).
%%CITATION = ASTRO-PH 0501317;%%

\bibitem{pao_topten}
Pierre Auger Collaboration,
Proc. 29$^{\rm th}$ Intern.~Cosmic Ray Conf., Pune, 
{\bf 7}, 283 (2005).
%%CITATION = FERMILAB-CONF-05-276-E-TD;%%

\bibitem{agasa_spectrum}
M.~Takeda {\it et al.}, Astropart.~Phys. {\bf 19}, 447 (2003).
%%CITATION = APHYE,19,447;%%

\bibitem{td_sigl}
P.~Bhattacharjee, G.~Sigl, Phys.~Rep.~{\bf 327}, 109 (2000).
%%CITATION = PRPLC,327,109;%%

\bibitem{sarkar}
S.~Sarkar, Acta Phys.~Polon.~{\bf B35}, 351 (2004).
%%CITATION = HEP-PH 0312223;%%

\bibitem{shdm}
V.~Berezinsky, M.~Kachelrie{\ss}, A.~Vilenkin,
Phys.~Rev.~Lett.~{\bf 79}, 4302 (1997);
%%CITATION = PRLTA,79,4302;%%
M.~Birkel, S.~Sarkar, Astropart.~Phys.~{\bf 9}, 297 (1998);
%%CITATION = APHYE,9,297;%%
S.~Sarkar, R.~Toldra,
%``The high energy cosmic ray spectrum from massive particle decay,''
Nucl.\ Phys.\  B {\bf 621}, 495 (2002);
%[arXiv:hep-ph/0108098].
%%CITATION = NUPHA,B621,495;%%
C.~Barbot, M.~Drees,
%``Detailed analysis of the decay spectrum of a super-heavy X particle,''
Astropart.\ Phys.\  {\bf 20}, 5 (2003);
%[arXiv:hep-ph/0211406].
%%CITATION = APHYE,20,5;%%
R. Aloisio, V. Berezinsky, M. Kachelriess,
Phys.~Rev.~D~{\bf 74}, 023516 (2006).
%%CITATION = PHRVA,D74,023516;%%

\bibitem{aloisio04}
R. Aloisio, V. Berezinsky, M. Kachelriess,
Phys.~Rev.~D~{\bf 69}, 094023 (2004).
%%CITATION = PHRVA,D69,094023;%%

\bibitem{ellis}
J.~Ellis, V.~Mayes, D.V.~Nanopoulos, Phys.~Rev.~D {\bf 74}, 115003
(2006).
%%CITATION = PHRVA,D74,115003;%%

\bibitem{td}
C.T.~Hill, Nucl.~Phys.~{\bf B224}, 469 (1983);
%%CITATION = NUPHA,B224,469;%%
M.B.~Hindmarsh, T.W.B.~Kibble, Rep.~Prog.~Phys.~{\bf 58}, 477 (1995).
%%CITATION = RPPHA,58,477;%%

\bibitem{zb}
T.J.~Weiler, Phys.~Rev.~Lett.~{\bf 49}, 234 (1982);
%%CITATION = PRLTA,49,234;%%
T.J.~Weiler, Astropart.~Phys.~{\bf 11}, 303 (1999);
%%CITATION = APHYE,11,303;%%
D.~Fargion, B.~Mele, A.~Salis, Astrophys.~J.~{\bf 517}, 725 (1999).
%%CITATION = ASJOA,517,725;%%

\bibitem{Roth_2007in}
Pierre Auger Collaboration,
%``Measurement of the UHECR energy spectrum using data from the
% Surface Detector of the Pierre Auger Observatory,''
Proc. 30$^{\rm th}$ Intern.~Cosmic Ray Conf., Merida (2007);
[arXiv:astro-ph/0706.2096].
%%CITATION = ARXIV:0706.2096;%%

\bibitem{busca}
N. Busca, D. Hooper, E.W. Kolb, Phys.~Rev.~D {\bf 73}, 123001 (2006).
%%CITATION = PHRVA,D73,123001;%%

\bibitem{desouza}
V.~de Souza, G.~Medina-Tanco, J.A. Ortiz, Phys.~Rev.~D~{\bf 72},
103009 (2005).
%%CITATION = PHRVA,D72,103009;%%

\bibitem{chou}
A.S. Chou, Phys.~Rev.~D {\bf 74}, 103001 (2006).
%%CITATION = PHRVA,D74,103001;%%

\bibitem{fd_limit}
Pierre Auger Collaboration,
Astropart.~Phys. {\bf 27}, 155 (2007);
%\cite{Healy:2007ef}
%\bibitem{Healy:2007ef}
%%CITATION = APHYE,27,155;%%
Pierre Auger Collaboration,
%``Search for Ultra-High Energy Photons with the Pierre Auger Observatory,''
Proc. 30$^{\rm th}$ Intern.~Cosmic Ray Conf., Merida (2007),
arXiv:0710.0025 [astro-ph].
%%CITATION = ARXIV:0710.0025;%%

\bibitem{gzk}
K.~Greisen, Phys.~Rev.~Lett.~{\bf 16}, 748 (1966);
%%CITATION = PRLTA,16,748;%%
G.T.~Zatsepin, V.A.~Kuzmin, JETP Lett.~{\bf 4}, 78 (1966).
%%CITATION = ZFPRA,4,114;%%

\bibitem{gelmini}
G.~Gelmini, O.E.~Kalashev, D.V.~Semikoz, [arXiv:astro-ph/0506128].
%%CITATION = ASTRO-PH/0506128;%%

\bibitem{sigl95}
S. Lee, A.V. Olinto, G. Sigl, Astrophys. J. {\bf 455}, L21 (1995).
%%CITATION = ASJOA,455,L21;%%

\bibitem{gzk-photon1}
G.~Sigl, Phys.~Rev.~D {\bf 75}, 103001 (2007) [arXiv:astro-ph/0703403].
%%CITATION = PHRVA,D75,103001;%%

\bibitem{gzk-photon2}
G.~Gelmini, O.~Kalashev, D.V.~Semikoz, [arXiv:astro-ph/0706.2181].
%%CITATION = ARXIV:0706.2181;%%

\bibitem{review}
M. Risse, P. Homola, Mod. Phys. Lett. A {\bf 22}, 749 (2007).
%%CITATION = MPLAE,A22,749;%%

\bibitem{galaverni}
M.~Galaverni, G.~Sigl, to appear in Phys.~Rev.~Lett.,
%``Lorentz Violation in the Photon Sector and Ultra-High Energy Cosmic Rays,''
[arXiv:astro-ph/0708.1737];
%%CITATION = ARXIV:0708.1737;%%
R.~Aloisio, P.~Blasi, P.~L.~Ghia, A.~F.~Grillo,
%``Probing the structure of space-time with cosmic rays,''
Phys.~Rev.~D {\bf 62}, 053010 (2000)
[arXiv:astro-ph/0001258].
%%CITATION = PHRVA,D62,053010;%%

\bibitem{lpm}
L.D.~Landau, I.Ya.~Pomeranchuk,
Dokl. Akad. Nauk SSSR {\bf 92}, 535 \& 735 (1953);
%%CITATION = DANKA,92,535;%%
%%CITATION = DANKA,92,735;%%
A.B.~Migdal, Phys. Rev. {\bf 103}, 1811 (1956).
%%CITATION = PHRVA,103,1811;%%

\bibitem{hp_photon}
M.~Ave {\it et al.}, Phys.~Rev.~Lett.~{\bf 85}, 2244 (2000);
%%CITATION = ASTRO-PH 0007386;%%
Phys.~Rev.~{\bf D65}, 063007 (2002).
%%CITATION = ASTRO-PH 0110613;%%

\bibitem{agasa_photon} 
K.~Shinozaki {\it et al.}, Astrophys.~J.~{\bf 571}, L117 (2002).
%%CITATION = ASJOA,571,L117;%%

\bibitem{risse_photon}
M.~Risse {\it et al.}, Phys.~Rev.~Lett. {\bf 95}, 171102 (2005).
%%CITATION = ASTRO-PH 0502418;%%

\bibitem{ay_limit}
G.I.~Rubtsov {\it et al.}, Phys.~Rev.~D~{\bf 73}, 063009 (2006).
%%CITATION = PHRVA,D73,063009;%%

\bibitem{yakutsk}
A.V. Glushkov et al., JETP Lett.~{\bf 85}, 163 (2007).
%%CITATION = JTPLA,85,131;%%

\bibitem{auger} 
%J.~Abraham {\it et al.}, P.~Auger Collaboration,
Pierre Auger Collaboration,
Nucl.~Instrum.~Meth.~{\bf A 523}, 50 (2004).
%%CITATION = NUIMA,A523,50;%%

\bibitem{sd_2007}
% Tiina S.
Pierre Auger Collaboration,
%``Performance of the Pierre Auger Observatory Surface Detector,''
Proc. 30$^{\rm th}$ Intern.~Cosmic Ray Conf., Merida (2007);
arXiv:0709.1823 [astro-ph].
%%CITATION = ARXIV:0709.1823;%%

\bibitem{bertou_photon}
X.~Bertou, P.~Billoir, S.~Dagoret-Campagne,
Astropart.~Phys. {\bf 14}, 121 (2000).

\bibitem{sdcal}
X.~Bertou {\it et al.}, P.~Auger Collaboration,
Nucl.~Instrum.~Meth.~{\bf A 568}, 839 (2006).
%%CITATION = NUIMA,A568,839;%%

\bibitem{icrc_angle} 
Pierre Auger Collaboration,
Proc. 29$^{\rm th}$ Intern.~Cosmic Ray Conf., Pune, {\bf 7}, 17 (2005).
%%CITATION = FERMILAB-CONF-05-301-E-TD;%%

\bibitem{s1000}
D.~Newton, J.~Knapp, A.A.~Watson, Astropart.~Phys. {\bf 26}, 414 (2007).
%%CITATION = APHYE,26,414;%%

\bibitem{sommers}
% P.~Sommers for the P.~Auger Collaboration,
Pierre Auger Collaboration,
Proc. 29$^{\rm th}$ Intern.~Cosmic Ray Conf., Pune,
{\bf 7}, 387 (2005), [arXiv:astro-ph/0507150].
%%CITATION = ASTRO-PH/0507150;%%

\bibitem{photon_energy}
P.~Billoir, C.~Roucelle, J.C.~Hamilton, [arXiv:astro-ph/0701583].
%%CITATION = ASTRO-PH/0701583;%%

\bibitem{icrc_hybrid}
%\cite{Dawson:2007di}
Pierre Auger Collaboration,
%``Hybrid Performance of the Pierre Auger Observatory,''
Proc. 30$^{\rm th}$ Intern.~Cosmic Ray Conf., Merida,
arXiv:0706.1105 [astro-ph].
%%CITATION = ARXIV:0706.1105;%%

\bibitem{icrc_uni}
A.S.~Chou {\it et al.},
Proc. 29$^{\rm th}$ Intern.~Cosmic Ray Conf., Pune, 
{\bf 7}, 319 (2005).
%%CITATION = FERMILAB-CONF-05-294-E;%%

\bibitem{preshower} 
T.~Erber, Rev.~Mod.~Phys.~{\bf 38}, 626 (1966);
%%CITATION = RMPHA,38,626;%%
B.~McBreen, C.J.~Lambert, Phys.~Rev.~D {\bf 24}, 2536 (1981);
%%CITATION = PHRVA,D24,2536;%%
T.~Stanev, H.P.~Vankov, Phys.~Rev.~D {\bf 55}, 1365 (1997);
%%CITATION = ASTRO-PH 9607011;%%
P.~Homola {\it et al.}, Comput.~Phys.~Commun. {\bf 173}, 71 (2005);
%%CITATION = ASTRO-PH 0311442;%%
P.~Homola {\it et al.}, Astropart.~Phys.~{\bf 27}, 174 (2007).
%%CITATION = APHYE,27,174;%%

\bibitem{klein}
S.~Klein, Rev.~Mod.~Phys.\ {\bf 71}, 1501 (1999);
% hep-ph/9802442
S.~Klein, Rad.~Phys.~Chem.\ {\bf 75}, 696 (2006).
% hep-ex/0402028

\bibitem{aires}
%S.J.~Sciutto, Departamento de Fisica Universidad Nacional de la Plata,
%Available from http://www.fisica.unlp.edu.ar/auger/aires/.
S.J.~Sciutto, ``AIRES: A system for air shower
simulations (version 2.2.0),''
arXiv:astro-ph/9911331;
available at  http://www.fisica.unlp.edu.ar/auger/aires
%Available from http://www.fisica.unlp.edu.ar/auger/aires/.
%%CITATION = ASTRO-PH/9911331;%%


\bibitem{qgsjet}
N.N.~Kalmykov, S.S.~Ostapchenko, A.I.~Pavlov,
Nucl. Phys. B (Proc. Suppl.) {\bf 52B}, 17 (1997).
%%CITATION = NUPHZ,52B,17;%%

\bibitem{geant4}
S.~Agostinelli {\it et al.}, Nucl.~Instrum.~Meth.~{\bf A 506}, 250
(2003).
%%CITATION = NUIMA,A506,250;%%
J.~Allison {\it et al.}, IEEE Transactions on Nuclear Science, {\bf
  53} 270 (2006).
See also http://geant4.cern.ch/.

\bibitem{icrc_ldf}
Pierre Auger Collaboration,
Proc. 29$^{\rm th}$ Intern.~Cosmic Ray Conf., Pune,
{\bf 7}, 291 (2005).
%%CITATION = ASTRO-PH/0507590;%%

\bibitem{sibyll}
%J.~Engel, T.K.~Gaisser, P.~Lipari, T.~Stanev,
%Phys.~Rev.~D {\bf 46}, 5013 (1992);
%R.S.~Fletcher, T.K.~Gaisser, P.~Lipari, T.~Stanev,
%Phys.~Rev.~D {\bf 50}, 5710 (1994);
R.~Engel, T.K.~Gaisser, P.~Lipari, T.~Stanev,
Proc. 26$^{\rm th}$ Intern.~Cosmic Ray Conf., Salt Lake City, 
{\bf 1}, 415 (1999).

\bibitem{inclined}
%  D.~Newton  [Pierre Auger Collaboration],
%``Selection and reconstruction of very inclined air showers with the Surface
%Detector of the Pierre Auger Observatory,''
Pierre Auger Collaboration,
Proc. 30$^{\rm th}$ Intern.~Cosmic Ray Conf., Merida (2007);
[arXiv:astro-ph/0706.2096].
%%CITATION = ARXIV:0706.3796;%%

\bibitem{pca_oja}
E.~Oja, International Journal of Neural Systems, {\bf 1}, 61 (1989).

\bibitem{pdg}
S.~Eidelmann {\it et al.},
Particle Data Group, Phys.~Lett.~{\bf B592}, 1 (2004).

\bibitem{risse_c2cr}
M.~Risse {\it et al.}, Czech.~J.~Phys. {\bf 56}, A327 (2006).
% [arXiv:astro-ph/0512434].
%%CITATION = ASTRO-PH 0512434;%%

\bibitem{rogers}
T.C.~Rogers, M.I.~Strikman, J.~Phys.~G: Nucl. Part. Phys. {\bf 32}, 2041 (2006).
%%CITATION = HEP-PH 0512311;%%

\bibitem{corsika}
D.~Heck {\it et al.}, Reports {\bf FZKA 6019 \& 6097},
Forschungszentrum Karls\-ruhe (1998).
%%CITATION = FZKA-6019;%%
%%CITATION = FZKA-6097;%%

\bibitem{semikoz_auger}
Pierre Auger Collaboration,
  %``Constraints on top-down models for the origin of UHECRs from the Pierre
  %Auger Observatory data,''
Proc. 30$^{\rm th}$ Intern.~Cosmic Ray Conf., Merida (2007);
  arXiv:0706.2960 [astro-ph].
%%CITATION = ARXIV:0706.2960;%%

\bibitem{science}
Pierre Auger Collaboration,
Science {\bf 318}, 939 (2007).

\end{thebibliography}
\end{document}